\documentclass[a4paper,fleqn,usenatbib]{mnras}

\usepackage[T1]{fontenc}
\usepackage{ae,aecompl}
\usepackage{relsize} 
\usepackage{color,soul}
\usepackage{subfig}

\usepackage{graphicx}	
\usepackage{amsmath}	
\usepackage{amssymb}	
\usepackage{bm}
\usepackage{footnote}

\title[External photoevaporation of PPDs in Cygnus OB2]{{External photoevaporation of protoplanetary discs in Cygnus OB2: linking discs to star formation dynamical history}}

\author[Winter, Clarke \& Rosotti]{Andrew~J.~Winter,\thanks{ajwinter@ast.cam.ac.uk} Cathie~J.~Clarke  and Giovanni~P.~Rosotti
\\
Institute of Astronomy, Madingley Road, Cambridge CB3 0HA, UK 
}

\date{Accepted XXX. Received YYY; in original form ZZZ}

\pubyear{2018}

\begin{document}
\label{firstpage}
\pagerange{\pageref{firstpage}--\pageref{lastpage}}
\maketitle

\begin{abstract}
Many stars form in regions of enhanced stellar density, where stellar neighbours can have a strong influence on a protoplanetary disc (PPD) population. In particular, far ultraviolet (FUV) flux from massive stars drives thermal winds from the outer edge of PPDs, accelerating disc destruction.  Here, we present a novel technique for constraining the dynamical history of a star forming environment using PPD properties in a strongly FUV irradiated environment. Applying recent models for FUV induced mass loss rates to the PPD population of Cygnus OB2, we constrain the time since primordial gas expulsion. This is $0.5$~Myr ago if the \citeauthor{Sha73} $\alpha$-viscosity parameter is $\alpha = 10^{-2}$ (corresponding to a viscous timescale of  $\tau_\mathrm{visc} \approx 0.5$~Myr for a disc of scale radius $40$~au around a $1\, M_\odot$ star). {This value of $\alpha$ is effectively an upper limit, since it assumes efficient extinction of FUV photons throughout the embedded phase.} This gas expulsion timescale is consistent with a full dynamical model that fits kinematic and morphological data as well as disc fractions. We suggest Cygnus OB2 was originally composed of distinct massive clumps or filaments, each with a stellar mass $\sim 10^4 \, M_\odot$. Finally we predict that in regions of efficient FUV induced mass loss, disc mass $M_\mathrm{disc}$ as a function of stellar host mass $m_\mathrm{star}$ follows a power law with $M_\mathrm{disc} \propto m_\mathrm{star}^\beta$, where $\beta$ exceeds $\sim 2.7$ -- steeper than correlations observed in regions of moderate FUV flux ($1 < \beta < 1.9$). This difference offers a promising diagnostic of the influence of external photoevaporation in a given region.
\end{abstract}




\section{Introduction}

The majority of stars form and spend their early stages of evolution in regions of enhanced stellar density \citep{Lad03}. This might be as part of a gravitationally bound cluster, or if star formation efficiency is sufficiently low this environment can become a short-lived association \citep{Lad03, Mur11,  Pfa13}. This latter case is expected to be common \citep[although see][]{Kru12}, therefore understanding the effect of clustered environments on the early stages of stellar evolution is important even for apparently isolated stars. 

In particular, feedback from the stellar environment has been shown to have a significant impact on planet formation in young clusters and associations. Protoplanetary disc (PPD) lifetimes are thought to be $\sim 3- 10$~Myr \citep[e.g][]{Hai01b, Wil11, Rib15}, during which period a star typically remains in its formation environment. In this case, PPD destruction can be induced due to close star-disc encounters \citep{Cla93, Ost94, Pfa05, Bre14, Win18, Vin18, Cue18, Con19} or far-ultraviolet (FUV) driven external photoevaporation \citep[and, to a lesser extent, extreme ultraviolet - EUV, ][]{Joh98, Sto99, Arm00, Ada10, Fac16, Haw18, Win18b}.  {Recent studies suggest that, statistically speaking, the dominant disc dispersal mechanism is external photoevaporation} \citep{Sca01, Win18b}, {while encounters can set PPD initial conditions during the evolution of stellar multiple systems} \citep{Win18c, Bat18}. 

FUV radiation has a significant influence on disc evolution in a wide range of environments. Mass loss from the outer edge is driven when the thermal energy exceeds the gravitational potential \citep[e.g.][]{Joh98}. The most obvious examples of such a process are proplyds, visible for example in the Orion Nebula Cluster \citep{Ode94}, {where the finite penetration of ionising radiation into a disc's neutral wind creates an offset (cometary) ionisation front. The strong neutral winds in the Orion Nebula Cluster are driven} by FUV flux of $F_\mathrm{FUV} \sim 3 \times 10^4 \, G_0$ \footnote{$G_0$ is the Habing unit \citep{Hab68}. It is a measure of the FUV field as a multiple of the solar neighbourhood value: $1.6 \times 10^{-3}$~erg~cm$^{-2}$~s$^{-1}$ .}. Since then proplyds have been found for discs which experience a factor $\gtrsim 10$ lower flux \citep{Kim16}. Even when a proplyd is not visible, a disc can exhibit significant FUV induced mass loss. \citet{Haw17b} find that in the very extended disc around IM Lup, with $F_\mathrm{FUV} \sim 4 \, G_0$,  photoevaporation drives substantial mass loss. A large fraction of stars form in regions where FUV flux is considerably greater than this \citep{Fat08, Win18b}.

The properties of young massive stellar clusters/associations and the giant molecular clouds (GMCs) from which they form are diverse, and the link between them is not well characterised \citep[see][for a review]{Lon14}. During formation the early cluster may undergo cold collapse \citep{Tob09, Kuz18}, or after the expulsion of gas the stellar population may become supervirial \citep{Goo09, Pfa13}, dependent on the density and velocity dispersion of the primordial GMC. This in turn influences the evolution of mass segregation \mbox{\citep[e.g.][]{Bon98}} and substructure \mbox{\citep[e.g][]{Goo04}} within the cluster. Because the environment of a star has an influence on the associated PPD, the dynamical history of a cluster is closely linked with the properties of a disc population.

While many authors have attempted to account for the local environment in considering PPD evolution {\citep[e.g.][]{Sca01,Cle16, Gua16}}, none have inverted this method and used observed disc populations to put constraints on the dynamical history of a star forming environment. This work is partly motivated by this goal. 

Cygnus OB2 (Cyg OB2) is a young massive OB association in the Cygnus X region, and has been used as an empirical test of feedback mechanisms on PPD evolution. It contains many massive stars up to $\sim 100 \, M_\odot$ \citep[e.g.][]{Mas91, Wri15} which contribute to strong FUV radiation fields. \citet{Gua16} analysed the disc fraction within Cygnus OB2 as a function of FUV flux, and found that surviving discs were less common at small projected separations from massive stars. Other authors, such as \citet{Wri16}, have made observations which indicate a complex dynamical substructure within the association. Collating this evidence, we here aim to apply $N$-body simulations and those combining viscous disc evolution and photoevaporation to replicate observations of Cyg OB2. We will reproduce the present day stellar kinematics and a dynamical history consistent with the observed disc fraction distribution. In this way we can shed light on both the history and the likely future of the PPD population and the stellar components. 

In the remainder of this work we first review the observational constraints on the properties of Cyg OB2 in Section \ref{sec:obsprops}. We describe our numerical method and models in Section \ref{sec:nummethod}. In Section \ref{sec:results} we compare our models with the observational data. We draw conclusions in Section \ref{sec:concs}.

\section{Properties of Cygnus OB2}
\label{sec:obsprops}

\subsection{Stellar population}
\label{sec:stellpop}
Cyg OB2 is a young association at a distance $\sim 1.33$~kpc from the Sun \citep{Kim15}. The majority of members formed $3-5$~Myr ago \citep{Wri10}, although some stars have ages as young as $\sim 2$~Myr \citep{Han03} and as old as $\sim 7$~Myr \citep{Dre08}. Estimates of the total stellar mass of Cyg OB2 have varied between $2$--$10 \times 10^4 \, M_\odot$ \citep{Kno00, Han03, Dre08, Wri10}, although \citet{Wri15} find a slightly lower mass of $\sim 1.6 \times 10^4 \, M_\odot$ within a radius of $13$~pc of the apparent centre. This includes a population of $\sim 169$ OB stars, the most massive of which is $\sim 100 M_\odot$ {with an age of $\sim 2$~Myr}. Cyg OB2 does not exhibit evidence of mass segregation, whereby the most massive stars occupy regions with the greatest gravitational potential \citep{Wri14}. 

The initial mass function (IMF) at the high mass end is not agreed upon in the literature, with many authors arriving at different conclusions \citep{Mas91, Mas95, Kno00, Kim07, Wri10, Com12}. \citet{Wri15} take into account massive stars which have evolved to their end state. In this way they find that the observed stellar masses follow an IMF $\zeta(m) \propto m^{-2.39\pm 0.19}$ at high masses, which is approximately consistent with the `universal' IMF of \citet{Kro01} \citep[or indeed a][IMF]{Sal55}. {However, inferring the high mass IMF is problematic since the occurance of supernovae in Cyg OB2 remains a point of debate} (see \citealt{But03} and discussion in  \citealt{Wri15}).

\citet{Wri16} describe the spatial density distribution in Cyg OB2 with an Elson, Fall and Freeman profile \citep[EFF profile hereafter - ][]{Els87}:
\begin{equation}
\label{eq:EFF}
\rho = \rho_0 \left( 1+ \frac{r^2}{a_\mathrm{stars}^2}\right)^{-\frac{\gamma+1}{2}}
\end{equation}
 where $a_\mathrm{stars} = 7.5$~pc and $\gamma=5.8$.  Normalising for the total mass of the cluster this makes the central mass density $\rho_0 \approx  22 \, M_\odot$~pc$^{-3}$. {However, this profile was derived for a small central region of $\sim 8$~pc$\times 8$~pc, which does not enclose the estimated core radius. Hence, when we consider the distribution of the stellar mass in our models} (see Section \ref{sec:centmass}) {we will focus on reproducing the mass enclosed within a projected radius of $13$~pc} \citep{Wri15}.

\subsection{Velocity dispersion}
\label{sec:obs_vdisp}
\citet{Wri16} presented a high-precision proper motion study {of the X-ray sources} in Cyg OB2. They found that the region is gravitationally unbound and exhibits an anisotropic velocity dispersion with proper motion components $\sigma_\alpha = 13.0 ^{+ 0.8}_{-0.7}$~km~s$^{-1}$ and $\sigma_\delta = 9.1 ^{+ 0.5}_{-0.5}$~km~s$^{-1}$. The radial velocity dispersion has also been measured to be $\sigma_r \sim 10$~km~s$^{-1}$, although systematic overestimates due to the binary fraction introduce uncertainties \citep{Kim07, Kim07err}.

Interestingly \citet{Wri16} found little evidence for expansion (or contraction) in the velocity field when considering the large scale variations of the proper motion distribution. This finding is independent of the definition of the centre of the association, and was argued by dividing proper motions into radial and azimuthal components. The ratio of kinetic energy between the radial and azimuthal directions was found to be approximately $60:40$. In the radial (projected) velocities, no bias is found towards or away from the centre. Azimuthally there is some preference for a ratio $66:34$ in favour of kinetic energy in the direction of negative PA. Because Cyg OB2 is not bound, this is interpreted by \citet{Wri16} as a remnant of the angular momentum of the primordial GMC. 

\subsection{Substructure}

Interpretation of the  internal substructure in Cyg OB2 is not straightforward. \citet{Kno00} concluded that it is a spherically symmetric region with a diameter of  $\sim 2^\circ$ ($46$~pc).  Since then, a number of authors have suggested a more complex morphology:

\begin{itemize}
	\item \citet{Bic03} suggested that Cyg OB2 is home to two open clusters which can be seen towards the centre. However, \citet{Gua13} note that the two apparent clusters are divided by a bright nebula. This makes it unclear if the two are physically separate or merely appear so due to the higher extinction in the intervening region. 
	\item \citet{Wri10} {found evidence of populations within Cyg OB2 with distinct ages, $3.5^{+0.8}_{-1.0}$~Myr and $5.25^{+1.5}_{-1.0}$~Myr for central and northwestern regions respectively. Ostensibly, this suggests multiple star forming events.} However, the ages in physically separated regions exhibited a wide spread such that they are almost consistent with being coeval. Further the authors acknowledge a number of sources of uncertainty, including variability in pre-MS stars \mbox{\citep{Her94}}; binarity \mbox{\citep{Pre99}}; variable accretion \citep{Bar09}; or non-uniform extinction \mbox{\citep{Gua12}}.
	\item A large number of A stars were identified south of the apparent centre by \citet{Dre08}. This population appears distinct spatially and non-coeval with the OB population (although the estimated ages $\sim 5$--$7$~Myr are nearly consistent with coevality within uncertainties). As the authors note, it is also possible that these stars are actually behind Cyg OB2 along the line-of-sight and wrongly associated due to projection \citep{Sch06}. 
	\item \citet{Gua13} used a critical side length criterion in the mimimum spanning tree of the disc-bearing population to suggest that Cyg OB2 has a clumpy substructure. However, as stated by the authors, the definitions of these clumps are dependent on the definition of the critical side length. Additionally the non-uniform extinction due to foreground gas complicates this argument as in the case of the two open clusters identified by \citet{Bic03}.
	\item Perhaps the best evidence for underlying structure in Cyg OB2 has been the proper motion study of \citet{Wri16}. On small scales they found evidence for kinematic substructure, which is the correlation of proper motion vectors with position. Applying a Moran's I statistic \citep{Mor50} they found correlation with a significance of $9.7\sigma$ and $12.5\sigma$ in RA and Dec velocity components respectively. 
\end{itemize}

\subsection{PPD population}

\citet{Gua16} studied the correlation between the fraction of surviving PPDs as a function of the local FUV and EUV intensity. {They use a sample of 7924 X-ray sources} \citep{Wri14b}, {for which the presence (absence) of a PPD is inferred by the presence (absence) of an infrared excess in the photometric data compiled from numerous surveys} \citep[see][]{Gua13}. Subsequently they estimate the local flux as a function of projected separation from each O star \citep[see][]{Gua07}. The disc fractions as a function of $F_\mathrm{FUV}$ are divided into 6 bins between $\sim 10^3 \, G_0$ and $\sim 5 \times 10^4 \, G_0$. Over this space the disc fraction drops monotonically from $\sim 40 \%$ to $18 \%$ with FUV intensity \citep[see figure 3 in][]{Gua16}. These observations are discussed in the context of modelling in Section \ref{sec:df_res1}.

\subsection{Observational summary \& modelling challenges}

Cyg OB2 is a well studied young association, and as such a large number of physical characteristics serve as constraints and measures for the success of any modelling attempts. Some such metrics are as follows:

\begin{itemize}
	\item \citet{Wri16} find that the velocity dispersion in Cyg OB2 is anisotropic. This suggests that the stars share the systematic large-scale velocity field of the primordial GMC. {We will find that such observations cannot be reproduced by simple models without underlying substructure, and we consider the initial properties required in Section} \ref{sec:vdisp_res}.	
	\item Cyg OB2 presently has a central mass  of $\sim 1.6 \times 10^4 \, M_\odot$ within $13$~pc of the centre. Any dynamical model should match this central density after evolving for the age of the association ($\sim 3-5$~Myr). {The initial stellar mass required to maintain this central density is explored in Section} \ref{sec:centmass}.
	\item Although Cyg OB2 is gravitationally unbound with a large velocity dispersion, there is no bias (inwards or outwards) in the radial component of kinetic energy. This apparently suggests a lack of recent rapid expansion, despite the high velocities. {When we consider our final $N$-body model, we will explore what expansion metric we would `observe' (Section} \ref{sec:bfmod_expsub}). 
	\item Statistical measures of the proper motion distribution suggest kinematic substructure which is probably indicative of stars travelling together as small virialised groups \citep{Wri16}. {We consider evidence for kinematic substructure for our final $N$-body model in Section} \ref{sec:bfmod_expsub}.
 \item The disc fraction as a function of (projected) FUV field strength \citep{Gua16} provides a constraint on the length of time for which external photoevaporation has been an efficient mechanism in Cyg OB2. This allows us to put constraints on the gas expulsion timescale. {We perform these calculations to constrain the appropriate $N$-body model in Section} \ref{sec:df_res1}, {then revisit the PPD properties obtained from our final model in Section} \ref{sec:df_final}.
\end{itemize}

\section{Numerical Method}
\label{sec:nummethod}

{The goal of our models is to reproduce the observed dynamical properties of the stellar population and the observed fractions of surviving discs. The latter is achieved by tracking the FUV flux experienced by PPDs evolving within a given $N$-body model. In this section we discuss the modelling procedure applied to both the stellar dynamics} (Section \ref{sec:Nbodymethod}) and the disc evolution (Section \ref{sec:PPDmodel}). 

\subsection{Kinematic modelling}
\label{sec:Nbodymethod}
The dynamical evolution of the stellar population is calculated using \textsc{Nbody6++GPU} \citep{Wan15}. This is an MPI/GPU accelerated version of \textsc{Nbody6} \citep{Spu99, Aar03}, and has built-in routines to deal with the evolution of a stellar cluster within a (gas) potential. As the stellar components of Cyg OB2 are presently highly supervirial, the latter feature is necessary for an initially virialised state. 

\subsubsection{Gas potential}
\label{sec:gpot}
The complex nature of Cyg OB2 means that some simplifying assumptions are required for dynamical modelling. We first assume that the cluster was initially in virial equilibrium due to the contribution to the gravitational potential of the gas in the primordial association. This is achieved by invoking a Plummer potential (for numerical convenience) corresponding to a gas density profile:
$$
\rho_\mathrm{gas} = \frac{3 M_\mathrm{gas}}{4 \pi a_\mathrm{gas}} \left( 1+ \frac{r^2}{a_\mathrm{gas}^2} \right)^{{-5}/{2}} 
$$ where $M_\mathrm{gas}$ is the total gas mass, $r$ is the radial distance from the centre of the association, and $a_\mathrm{gas}$ is the scale parameter. {During initial tests, we have varied $a_\mathrm{gas}$ (and the stellar scale parameter $a_\mathrm{stars}$), and although it has a mild effect on the kinematic properties of our models, the exact value (within order unity) is not crucially important. {This is particularly true in the stochastically defined initial conditions of substructured models} (see Section \ref{sec:stellics} below), where kinematics are more dependent on the specific realisation. Physically we expect $a_\mathrm{gas} \gtrsim a_\mathrm{star}$, and we fix $a_\mathrm{star}=7$~pc (as there is no clear evidence of past expansion) and $a_\mathrm{gas}=10$~pc. 

For a Plummer density profile the specific gravitational potential is:
\begin{equation}
\label{eq:plummerpot}
\phi(r) = -\frac{GM_\mathrm{gas}}{a_\mathrm{gas}} \left(1 + \frac{r^2}{a_\mathrm{gas}^2} \right)^{-1/2} .
\end{equation} The total potential in a given cluster make up of $N$ stars of mass $m_i$ initially at distance $r_i$ from the centre and $r_{ij}$ from a star of mass $m_j$ is:
\begin{equation}
\label{eq:totalpot}
U_\mathrm{tot} = \sum_{i}^{N}  m_i \left(\phi(r_i) + \sum_{j \neq i}^N \frac{G m_j}{2 r_{ij}} \right).
\end{equation} Then using equations \ref{eq:plummerpot} and \ref{eq:totalpot} we require that initially 
$$
Q_{\mathrm{vir, }0} \equiv \frac{ \sum_{i}^N m_i v_i^2}{2U_\mathrm{tot}} = 0.5
$$ 
where $v_i$ is the magnitude of the initial velocity of the $i^\mathrm{th}$ star. The total gas mass is chosen to maintain initial virial equilibrium for a cluster with a given velocity dispersion. 

Our prescription of gas removal introduces an expulsion timescale $\tau_\mathrm{exp}$ over which time the potential is removed. The gas mass is reduced linearly such that $\dot{M}_\mathrm{gas} = M_{\mathrm{gas, }0}/\tau_\mathrm{exp}$. We vary $\tau_\mathrm{exp}$ to investigate how this affects the disc population due to extinction (see Section \ref{sec:df_res1}). We fix the time at which gas expulsion is initiated to be $\tau_\mathrm{delay} = 1$~Myr, consistent with the age of the most massive stars in Cyg OB2 ($\sim 2 $~Myr) if the cluster age is $3$~Myr (the period for which we evolve the whole system). We further define $\tau_\mathrm{gas} \equiv \tau_\mathrm{delay}+\tau_\mathrm{exp}$. For a discussion of the influence of gas expulsion on the dynamical state of a young cluster, see \citet{Bau07}.

{Clearly, while computationally necessary here, a Plummer potential is not a realistic reflection of the physical conditions of the primordial gas distribution during the embedded phase. Initially, gas density distributions would be physically expected to trace the stellar density (since the stars form from the gas). Subsequently we would expect gas expulsion to occur as expanding bubbles from the most massive stars in the region} \citep[e.g.][]{Dal11, Dal14, Ali18}.  {In terms of the cluster evolution, we therefore expect intra-clump or -filament potential to reduce faster than inter-clump potential. The influence of a geometrically complex potential on the evolution of the stellar population is certainly of interest for accurately reproducing observed kinematics and spatial distributions. However, we assume that the Plummer potential imposed here is sufficient for reproducing the global distribution of stellar positions and velocities. This is justified since we are primarily interested in preventing the rapid escape of the high velocity stars from the central regions of Cyg OB2. This is achieved by our spherically symmetric potential.}

\subsubsection{Stellar initial conditions}
\label{sec:stellics}

\begin{figure*}
     \subfloat[\label{subfig:frac} \texttt{FRAC}]{%
       \includegraphics[width=0.45\textwidth]{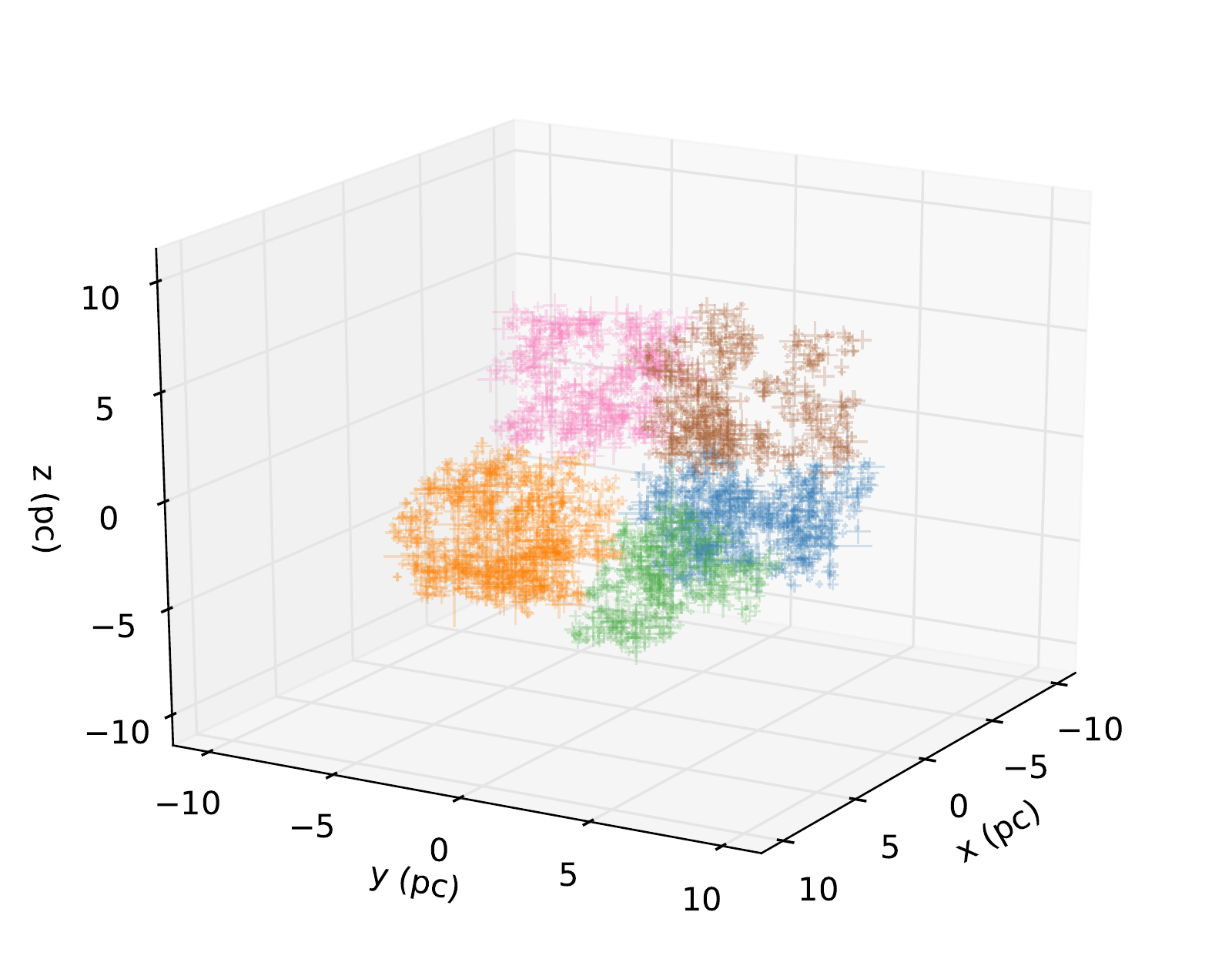}
     }
     \subfloat[\label{subfig:fila} \texttt{FILA}]{%
       \includegraphics[width=0.45\textwidth]{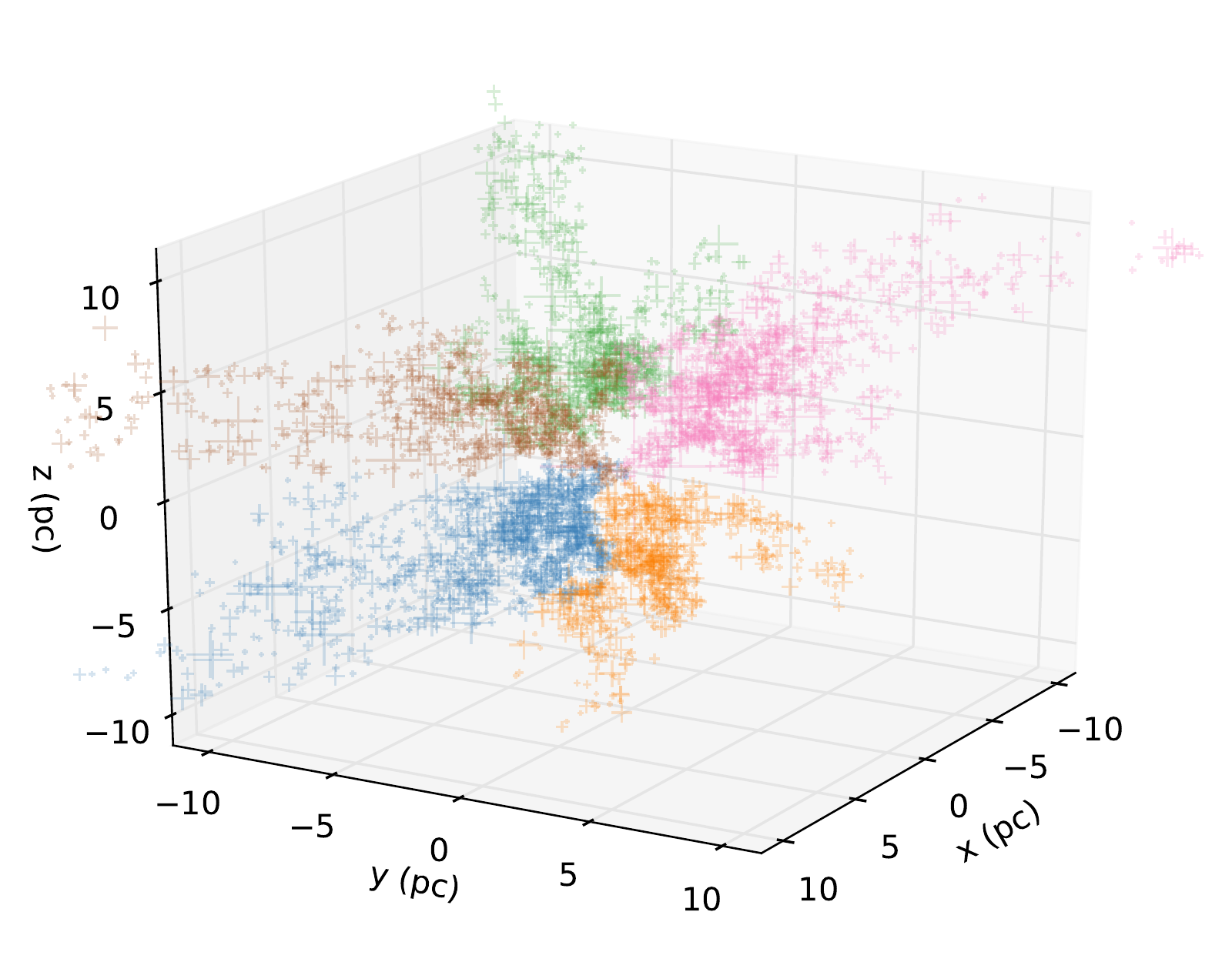}
     }
     \caption{Example initial position distribution of stars in \texttt{FRAC} and \texttt{FILA} models (figures \ref{subfig:frac} and \ref{subfig:fila} respectively). {In the \texttt{FRAC} model, stellar positions are distributed with a `clumpy' morphology, whereas the \texttt{FILA} model exhibits extended filaments. }A subset of $5000$ stars within a cube of side length $10$~pc are shown, where the coordinate system is defined by the gas potential (see text for details). Both models have a stellar mass of $10^4 \, M_\odot$ and the same half mass as an EFF profile with $a_\mathrm{stars}=7$~pc, $\gamma = 5.8$; the \texttt{FILA} model also follows the same radial density profile (equation \ref{eq:EFF} with the aforementioned parameters). The scatter points are coloured by the largest scale subgroup with which they are associated and scaled linearly by the mass of the star.  }
     \label{fig:ics}
   \end{figure*}
We define three different types of stellar initial condition, which we label uniform (\texttt{UNIF}), fractal (\texttt{FRAC}), and filamentary (\texttt{FILA}). A uniform cluster exhibits no underlying substructure, while fractal and filamentary clusters have enhanced local number densities, and stellar positions are correlated with velocities. {While the surface density of \texttt{FRAC} model is `clumpy', and individual clumps can be spatially isolated, a \texttt{FILA} model is defined such that the (radially binned) surface density follows an EFF profile. A filamentary model captures the morphology seen in both observations and simulations of star forming regions} \citep[e.g.][]{Bon08, Mol10, And14}, however we include both \texttt{FRAC} and \texttt{FILA} models in this work to compare their properties by the metrics of interest. We discuss the generation of each set of initial conditions below.  

The simplest initial conditions are \texttt{UNIF} models for which stellar positions are simply drawn from an EFF profile \citep[equation \ref{eq:EFF}, ][]{Els87}. We fix $\gamma = 5.8$, consistent with the present day value, and fix $a_\mathrm{stars} =7$~pc. The magnitude of the velocities are then chosen from a Maxwell-Boltzmann distribution
\begin{equation}
\label{eq:MBdist}
f(v) \propto v^2 \exp(-v^2)
\end{equation} with a random direction, then normalised as discussed in Section \ref{sec:gpot}. We note that drawing from this distribution (particularly at the high velocity dispersions we require - see Section \ref{sec:vdisp_res}) {will produce a significant number of unbound stars. It is possible that this distribution is truncated at large velocities, and indeed star formation models suggest stars might have a smaller velocity dispersion than the priomordial gas} \citep[][see Section \ref{sec:summary_bfm}]{Off09}.

A \texttt{FRAC} model is generated using the recipe fully described by \citet{Goo04} which we briefly review here \citep[see also][]{Sca02,Cra13}. First we define a cube with side length $2$ (in arbitrary units) centred at the origin. We then divide it up into $(2 P_0)^3$ sub-cubes, where $P_0$ is an integer (initially chosen to be unity) which dictates the number of largest scale sub-clusters. The centre of each of these represent the potential positions for the first generation of stars, $g=1$. All positions are offset by a vector with magnitude uniformly drawn between $0$ and $2^{-(g+1)}/P_0$, and random (isotropically drawn) direction. `Parent' positions $\bm{r}_{g}$ have $8$ possible sites for `child' positions $\bm{r}_{g+1}$ which are placed with a probability $2^{D_0-3}$, where $D_0 \leq 3$ is the fractal dimension (which we fix at $2.5$). Only existing children can parent future generations. We repeat this process until the number of positions greatly exceeds the number of stars in the model, at which point the members are randomly allocated to positions. The side length of the original cube is then redefined such that the initial half-mass radius matches that of an equivalent \texttt{UNIF} model with parameters $a_\mathrm{stars}=7$~pc, $\gamma=5.8$.

The velocities for each generation of stars $g=1$ is chosen in the same way as for the \texttt{UNIF} model. Velocities for subsequent generations $\bm{v}_{g+1}$ are correlated to the velocity of the parent star $\bm{v}_{g}$:
$$
\bm{v}_{g+1} = \bm{v}_{g} +\delta \bm{v}_{g+1}
$$ where $\delta \bm{v}_g$ is a velocity with a random direction and magnitude $\delta v_g$. The latter is drawn from the modified Maxwell-Boltzmann distribution
$$
f(\delta v) \propto \delta v^2 \exp \left( -2^{g} \delta v^2 \right)
$$ such that child stars have velocities correlated to their parents.

A \texttt{FILA} model is a hybrid between \texttt{FRAC} and \texttt{UNIF}. To construct the initial conditions we first produce a \texttt{FRAC} model and then force the stellar positions into an EFF profile. {This is achieved by dividing the radial positions into bins, with the $j^\mathrm{th}$ bin containing $N_{\mathrm{c},j}$ stars approximately at radius $r_{\mathrm{c},j}$, and rescaling the size of each bin ($\Delta r_{\mathrm{c},j}$) from inside out to produce the appropriate number density with respect to imposed EFF profile:}
$$
\frac{N_{\mathrm{c},j}}{4\pi r_{\mathrm{c},j}^2 \Delta r_{\mathrm{c},j}}  = \frac{\rho(r_{\mathrm{c},j})}{\langle m_\mathrm{star} \rangle}
$$ where $\langle m_\mathrm{star} \rangle$ is the average stellar mass and $\rho(r)$ is defined in equation \ref{eq:EFF}.

The results of this process are filament-like structures, as shown in figure \ref{fig:ics}, in which \texttt{FRAC} and \texttt{FILA} models are compared. In a \texttt{FRAC} model the density profile has hard edges and a clump-like substructure {(as in figure} \ref{subfig:frac}), while in the \texttt{FILA} model stellar density drops off smoothly with radius and produces filament-like substructure (as in figure \ref{subfig:fila}). {This is because the initial clumps from which the stellar density is composed become `stretched' radially when we impose the EFF profile, to produce several elongated distributions of stars.} Both types of model demonstrate spatial and kinematic asymmetry with respect to the gas potential.

We draw stellar masses from a \citet{Kro01} IMF: 
\begin{equation}
\label{eq:imf}
  \xi(m)\propto \begin{cases}
               m^{-1.3} \quad \mathrm{for } \, 0.08 \, M_\odot \leq m < 0.5 \, M_\odot\\ 
              m^{-2.3} \quad \mathrm{for } \, 0.5 \, M_\odot \leq m < 1.0 \, M_\odot\\
              m^{-2.4} \quad \mathrm{for } \, 1.0 \, M_\odot \leq m < 100 \,M_\odot \\
              0 \qquad \quad \, \mathrm{else}
            \end{cases}
\end{equation} where at high mass end $>1 \, M_\odot$ we use the observed mass function in Cyg OB2 \citep{Wri15}. In our models stars are not primordially mass segregated.

{Reproducing observations requires estimating an appropriate field of view, which is in turn dependent on the definition of the cluster centre. Observationally some discrepancy exists in this definition between different works, although authors generally agree within a few pc} \citep[see][for a discussion]{Wri16}. {For our purposes an approximate estimate of this centre is sufficient as we find that all of the metrics we consider are only weakly dependent on our choice. For \texttt{UNIF} models, the cluster centre remains the centre of mass of the original set up. For \texttt{FILA} and \texttt{FRAC} models, where considerable anistropy exists in the stellar kinematic distribution, it is necessary to estimate the centre of mass for each snapshot in time. We choose an efficient (approximate) algorithm, in which we find the point which maximises the mass within a given projected radius (chosen to be $R_\mathrm{cent} = 10$~pc) by sampling recursively over a grid of points. Providing the grid is sufficiently highly resolved, the centre we obtain is insensitive to the exact value of $R_\mathrm{cent}$ and number of iterations. All subsequent results should be understood in this context.}

\subsection{Disc evolution model}
\label{sec:PPDmodel}

{In this section we discuss the prescription we apply to calculate PPD evolution. Each disc is exposed to an FUV flux resulting from tracking the contributions from the stellar components within a given $N$-body model.}

\subsubsection{FUV flux and mass loss rate}

Flux in the FUV energy range ( $6 $~eV$< h\nu < 13.6$~eV\footnote{{Photons with $ h\nu > 13.6$~eV are considered extreme ultraviolet (EUV).}}) is the most efficient contributer to external photoevaporation of PPDs in an intermediate regime of distances from massive stars. As discussed in \citet{Win18b}, extreme ultraviolet radiation can also contribute to the thermal wind, however it only dominates the mass loss rate in regions where FUV flux is low ($F_\mathrm{FUV} \lesssim 50 \, G_0$) or very high ($F_\mathrm{FUV} \gtrsim 10^5\, G_0$). Both of these thresholds lie outside the region of interest in Cyg OB2. Additionally, in these cases disc destruction timescales are extremely long and short respectively, and the additional mass loss contributed by EUV flux in young massive clusters is therefore not significant to our consideration of the survival rates of discs on a timescale of Myr. 

To calculate the FUV luminosity for a star of a given stellar mass we apply the precription discussed by \citet{Arm00} and \citet{Win18b}. Model grids of \mbox{\citet{Sch92}} contain the total luminosities and effective temperatures of stars of a given mass, for which we use the results for metallicity $Z=0.02$ and the output age closest to $1$~Myr. We then obtain the wavelength dependent luminosity from the atmosphere models of \mbox{\citet{Cas04}}. In this way we obtain an FUV luminosity for all stars with a mass $m$ in the range $1\, M_\odot < m< 100 \, M_\odot$. 

For the FUV induced mass loss rates $\dot{M}_\mathrm{wind}$ we use the recent grid of models calculated by \citet{Haw18b}. The grid covers a wide range of parameter space in outer disc radius ($1$ -- $400$~au), disc masses ($\sim 10^{-8}$ -- $0.1  \, M_\odot$), FUV field strengths ($10$ -- $10^4$~$G_0$) and stellar masses ($0.05$ -- $1.9\, M_\odot$). These mass loss rates are interpolated linearly and applied to a viscously evolving disc to establish the expected disc properties in a given cluster environment. 

\subsubsection{Viscous disc evolution}
\label{sec:visc_evol}
\begin{figure}
       \includegraphics[width=0.5\textwidth]{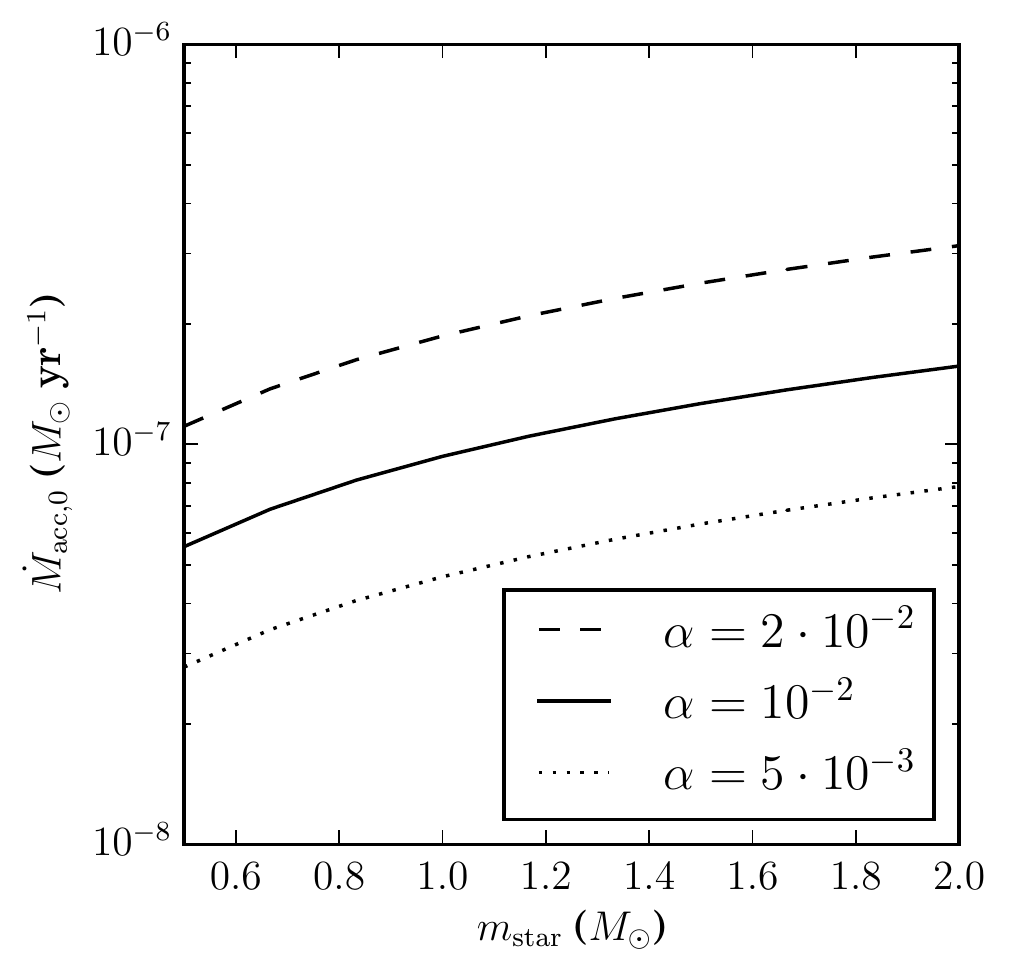}
     \caption{Assumed intial viscous accretion rate (equation \ref{eq:m_acc}) as a function of stellar mass for $M_{\mathrm{disc},0} = 0.1 \, m_\mathrm{star}$ and a range of \citeauthor{Sha73} $\alpha$-viscosity parameters. The initial disc conditions are described in Section \ref{sec:visc_evol}. The range of stellar masses we consider ($0.5$--$2$~$M_\odot$) is discussed in Section \ref{sec:df_res1}.}
     \label{fig:macc0}
   \end{figure}

To calculate the state of PPDs evolving within the cluster we must take into account vicous expansion as well as the photoevaporative mass loss. The viscous disc evolution is modelled using the method of \citet[][and subsequently \mbox{\citealt{And13}}, \mbox{\citealt{Ros17}} and \mbox{\citealt{Win18b}}]{Cla07}. In such a paramerisation, viscosity is assumed to scale linearly with radius $r$ within the disc, corresponding to a constant $\alpha$-viscosity parameter \citep{Sha73} and a temperature which scales with $r^{-1/2}$. {The accretion rate at the inner edge of the disc is initially}
\begin{equation}
\label{eq:m_acc}
\dot{M}_{\mathrm{acc},0} = \frac{3 \alpha \, M_{\mathrm{disc, }0}  H_1^2 \Omega_1}{2 R_1^2}
\end{equation} {where  $M_{\mathrm{disc, }0}$ is the initial disc mass, $H_1$ and $\Omega_1$ are the scale height and Keplerian frequency at the disc scale radius $R_1$ }\mbox{\citep[see][]{Har98}}. Providing that the outer radius of the disc $R_{\mathrm{disc},0}$ is significantly greater than $R_1$, the initial surface density can be written
\begin{equation}
\label{eq:surfdense}
\Sigma_0(r<R_{\mathrm{disc},0}) \approx \frac{ M_{\mathrm{disc, }0}}{2\pi R_1^2} \left( \frac{r}{R_1} \right)^{-1} e^{-r/R_1}
\end{equation} and $\Sigma_0(r>R_{\mathrm{disc},0}) =0$. We choose  $R_{\mathrm{disc},0} = 2.5 R_1$, such that the integral of equation \ref{eq:surfdense} over the disc area yields $92 \%$ of $M_{\mathrm{disc},0}$. We take the scale radius to be
\begin{equation}
\label{eq:R1}
R_1 = 40 \left( \frac{m_{\mathrm{star }}}{1 \, M_\odot}\right)^{1/2}  \, \mathrm{au}.
\end{equation} Unless otherwise stated we will assume that $M_{\mathrm{disc},0}$ is uniformly distributed between $0.01$ -- $0.1 \, m_\mathrm{star}$ \citep{Andr13}; then equation \ref{eq:R1} means that the distribution of initial surface densities at $R_1$ remains independent of stellar host mass. The scale height $H$ is proportional to the radius throughout the disc, and we choose $H_1/R_1 = 0.05$. The maximum initial accretion rate (equation \ref{eq:m_acc}) as a function of stellar mass is indicated in figure \ref{fig:macc0}. The associated viscous timescale is
\begin{equation}
\label{eq:tvisc}
\tau_\mathrm{visc}  \approx \frac{5.4 \cdot 10^3}{ \alpha} \left( \frac{R_1}{40 \, \mathrm{au}}\right)^{3/2}  \left( \frac{m_\mathrm{star}}{1 \,M_\odot}\right)^{-1/2} \, \mathrm{yr} 
\end{equation} or $\tau_\mathrm{visc} \approx 0.5$~Myr for a solar mass star with $\alpha = 10^{-2}$, $R_1 =40$~au. Equations \ref{eq:R1} and \ref{eq:tvisc} yield $\tau_\mathrm{visc} \propto m_\mathrm{star}^{1/4}$. 

During the course of this work we will explore the effect of altering disc initial conditions on their final properties, however we always consider a distribution of initial disc masses. {Allowing a range of initial disc masses accounts for the observed range of stellar ages (since internal processes deplete the disc over time)  and variable disc initial conditions.} Additionally, for a given FUV flux environment some fraction of discs survive, and the findings of \citet{Gua16} indicate that this survival fraction reduces monotonically with increasing $F_\mathrm{FUV}$. {We find that, depending on initial disc properties, dynamical mixing between regions of different FUV flux and the range of $F_\mathrm{FUV}$ within a single bin alone are insufficient to produce the observed survival fractions} {(i.e. observed survival fractions between $F_\mathrm{FUV}$ bins do not jump rapidly from $\sim 0 \%$ to $\sim 100 \%$ at a certain threshold)}. The chosen initial conditions and the variation between discs are therefore important in reproducing observations. Whether or not this dispersion is inherited from a tight correlation between stellar mass and PPD initial conditions is explored by considering host mass independent disc models in Section \ref{sec:mi_res}.

Using the initial conditions described above we calculate the evolution of the disc over a 1D grid, spaced evenly in $r^{1/2}$. The cell at the inner edge has a zero torque boundary condition, and the cell at the outer edge experiences mass depletion due to the photoevaporation induced wind (i.e. for a given mass loss rate $\dot{M}_\mathrm{wind}$, the disc material is removed from the outer edge). The edge cell is then redefined depending on whether there is net mass loss or mass accumulation \citep[see Section 4.2 in][]{Win18b}.

\section{Results and discussion}
\label{sec:results}
\subsection{Modelling approach}

{We aim to produce an $N$-body model, including initial gas potential, with a self-consistent treatment for the photoevaporation of the PPD population. We simplify the modelling procedure by the following approach:}
\begin{enumerate}
\item First we estimate the gas expulsion timescale $\tau_\mathrm{gas}\equiv  \tau_\mathrm{delay}+\tau_\mathrm{exp}$ by considering the observed surviving disc fractions as a function of FUV field strength (Section \ref{sec:df_res1}). The presence of primordial gas influences the models in two ways: it imposes a gravitational potential on the stellar population, and suppresses photoevaporation by extincting FUV photons. Using the latter effect we can calibrate the period of efficient exposure to the observed survival rate of PPDs. There exists a degeneracy in this calculation with the assumed disc viscosity, and we explore the influence of varying both parameters  (see Section \ref{sec:df_res1}).
\item Having established the rate of gas expulsion, we apply the appropriate time-dependent potential to establish the dynamical evolution of the stellar population consistent with kinematic and spatial data. In Section \ref{sec:vdisp_res} we deduce the initial conditions required to reproduce the observed anistropic velocity dispersion. In Section \ref{sec:centmass} we vary the initial stellar mass required to reproduce the observed central density (within $13$~pc of the apparent centre). 
\item Finally we combine our findings into a `best-fitting' model, and explore the evolution of PPDs over time in Section \ref{sec:df_final}. This allows us both to test whether external photoevaporation is  a viable mechanism for disc depletion in Cyg OB2 and to make predictions regarding the disc properties for future observations.
\end{enumerate}

\subsection{Disc fractions and gas expulsion}
\label{sec:df_res1}
\begin{figure*}
     \subfloat[\label{subfig:texp1} $\tau_\mathrm{gas} =1$~Myr ]{%
       \includegraphics[width=0.45\textwidth]{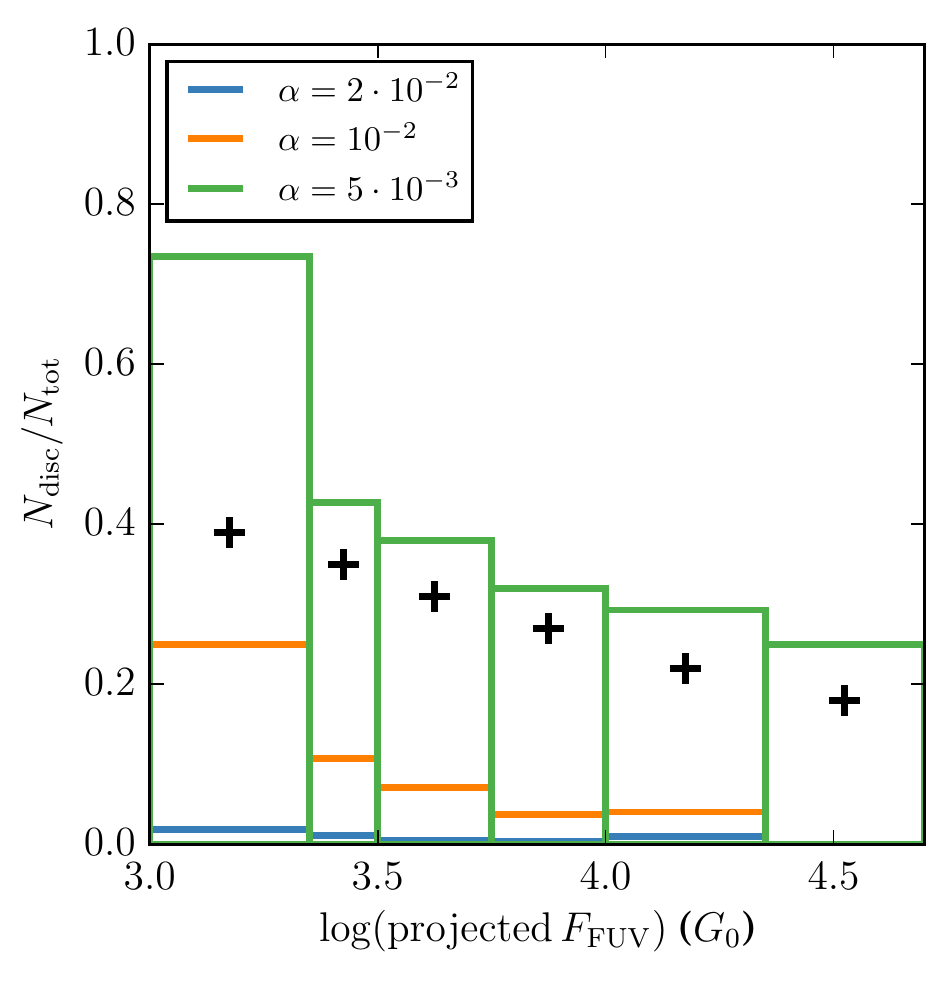}
     }
     \subfloat[\label{subfig:texp2}  $\tau_\mathrm{gas} =2$~Myr ]{%
       \includegraphics[width=0.45\textwidth]{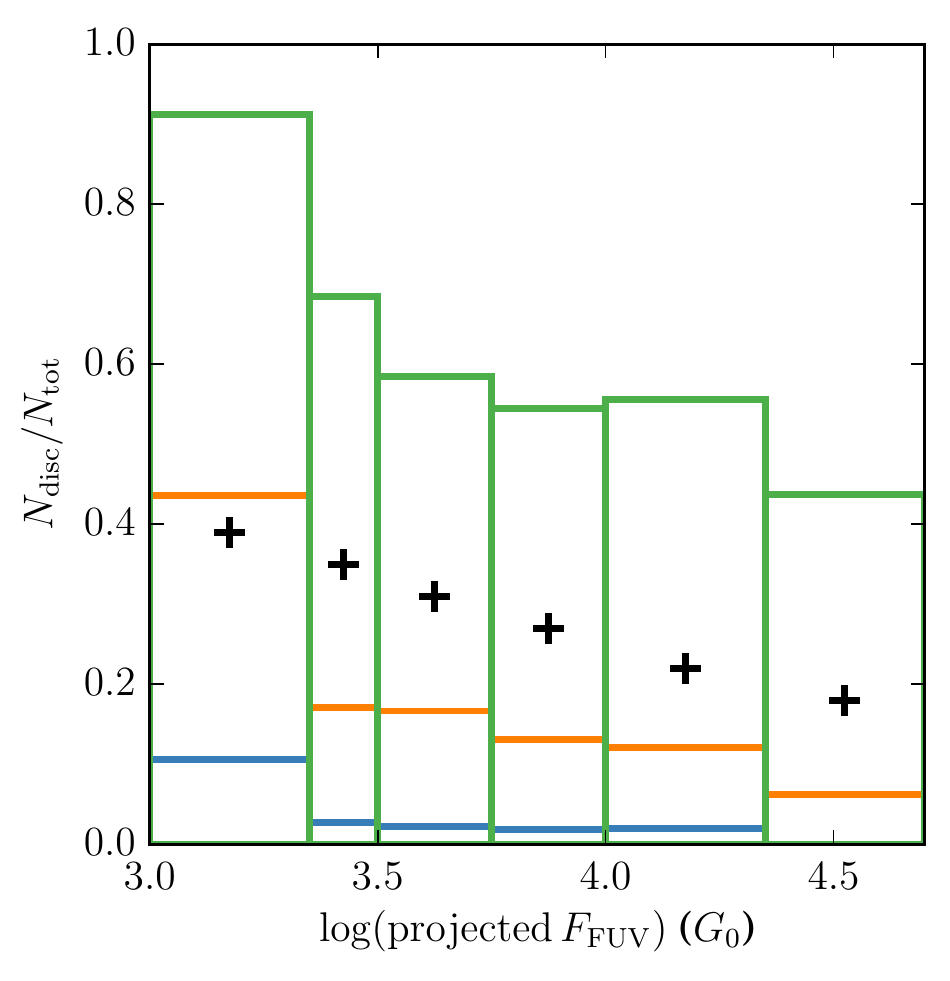}
     } \hfill
     \subfloat[\label{subfig:texp2.5} $\tau_\mathrm{gas} =2.5$~Myr ]{%
       \includegraphics[width=0.45\textwidth]{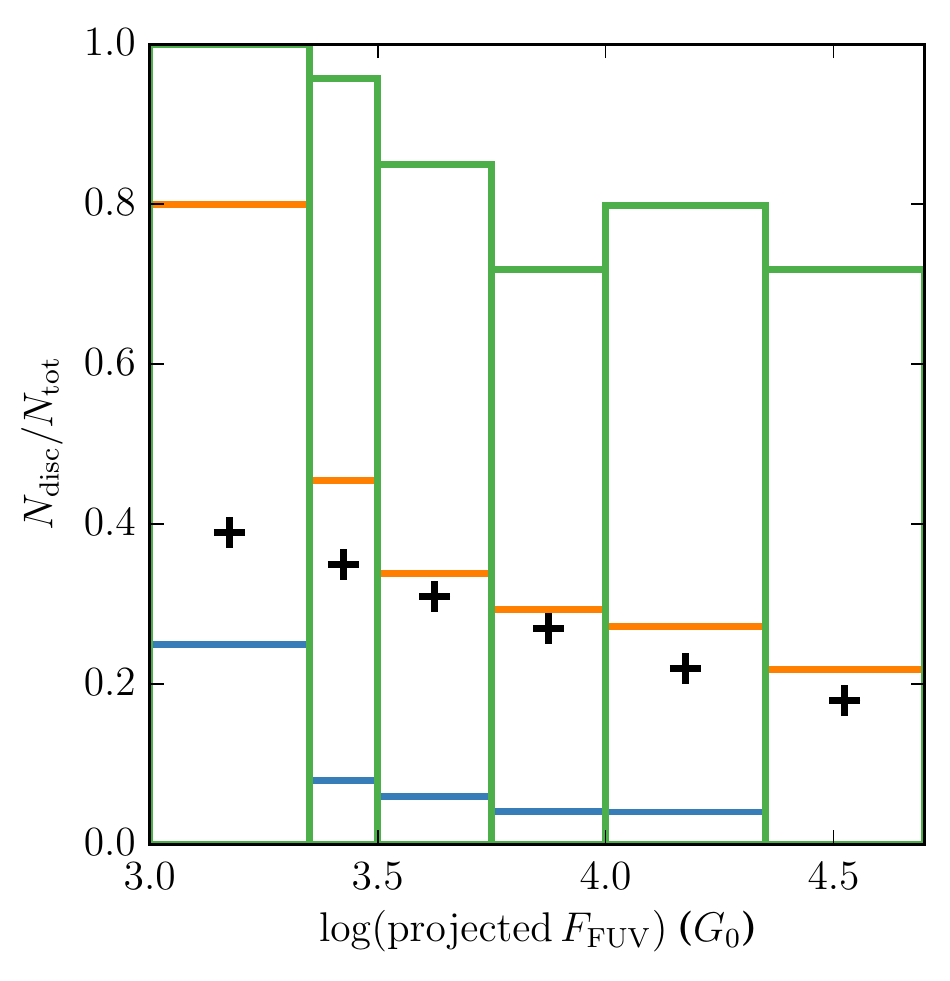}
     }
     \subfloat[\label{subfig:texp2.75} $\tau_\mathrm{gas} =2.75$~Myr ]{%
       \includegraphics[width=0.45\textwidth]{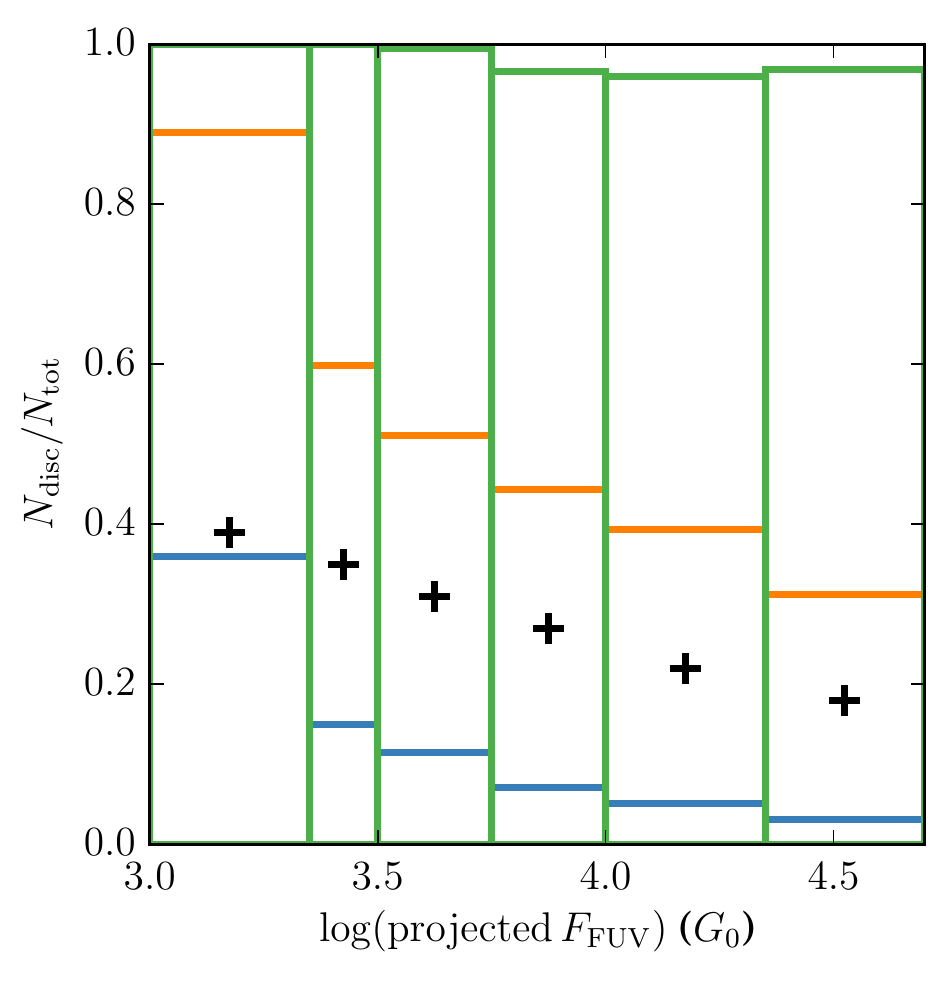}
     }
     \caption{Disc fractions versus projected FUV flux in a virialised cluster evolved for $3$~Myr when external photevaporation is `switched on' after a period $\tau_\mathrm{gas}$. Results are shown for a range of \citeauthor{Sha73} $\alpha$-viscosity values. The black crosses represent the observational values found by \citet{Gua16}. {These results are used to calibrate the timescale for gas expulsion and the corresponding disc viscosity required to reproduce the observed disc fractions. We find that $\tau_\mathrm{gas}=2.5$~Myr with $\alpha = 10^{-2}$ matches observed disc fraction. This value for $\alpha$ is effectively an upper limit since we assume that extinction efficiently shuts off photoevaporation before all gas is expelled.} }
     \label{fig:alpha_discfracs}
   \end{figure*}

{We first estimate the influence of the timescale for gas expulsion on the PPD population survival fractions.} A number of physical effects must be taken into account when considering the statistical distribution of disc fractions. Most obviously, projection effects can lead to a distribution of physical distances between stars for every apparent (projected) separation, and therefore a range of $F_\mathrm{FUV}$. Binarity and the initial PPD properties will also influence the total disc fraction. However, to first order, the steepness of the drop-off of the disc fraction with $F_\mathrm{FUV}$ indicates the length of time for which external photoevaporation has been an efficient mechanism for disc destruction in a given enivironment. Initial tests suggest that a massive gas mass ($\gtrsim 10^6 M_\odot$) is required to maintain virial equilibrium in the primordial Cyg OB2. Due to extinction of FUV photons, this gas mass is sufficient to dramatically reduce photoevaporation efficiency.

{The current PPD population therefore allows us to constrain when the gas component of Cyg OB2 was expelled. The relatively short period of expulsion ($\lesssim 2$~Myr) means that we are free to consider the influence of FUV photons on disc evolution from the time $\tau_\mathrm{gas}$ at which gas is completely removed.} Practically this means that we can apply a simplified \texttt{UNIF} model with the present day mass and `switch on' photoevaporation at different times. We then compare the disc fractions as a function of FUV flux after $3$~Myr of evolution, and thus estimate the gas expulsion timescale. The FUV flux in those bins is calculated in the same way as in \citet{Gua16}, using the projected distance between stars - this we call `projected' $F_{\mathrm{FUV}}$ (as opposed to `real' $F_\mathrm{FUV}$, as experienced by a given disc). As an estimate of the influence of dynamical mixing between projected FUV flux bins, we start with a stellar velocity dispersion $17$~km/s and hold the stars in virial equilibrium with an external potential.

The rate at which irradiated discs are destroyed is also dependent on the $\alpha$-viscosity parameter (Section \ref{sec:visc_evol}). {The chosen $\alpha$ dictates the rate at which a disc viscously expands into a region where photoevaporation rates are high, as well as dictating mass loss through accretion.} {This adds a degeneracy to our approach which, given uncertainties in $\alpha$, introduces similar uncertainties in $\tau_\mathrm{gas}$. We investigate this degeneracy by calculating surviving PPD fractions for a number of different values for $\tau_\mathrm{gas}$ ($1$, $2$, $2.5$ and $2.75$~Myr) and $\alpha$ ($5 \times 10^{-3}$, $10^{-2}$, $2\times 10^{-2}$). }

We calculate disc evolution for a subset of discs with host stars in the mass range $0.5$ -- $2$~$M_\odot$. Stars less massive than this are not present in the sample used by \citet{Gua16}, while the disc mass loss rates calculated by \citet{Haw18b} do not apply for higher mass stars. A disc is considered destroyed if it has a mass $< 10^{-5} \, M_\odot$, which is estimated by \mbox{\citet{Gua13}} as a limit below which SEDs are more difficult to interpret. In fact discs of such low mass ($\sim 10^{-5} \, M_\odot$) are destroyed quickly in regions of strong FUV fields, so our results are insensitive to the exact value of this threshold.

The results of this preliminary modelling procedure compared with the observational findings of \mbox{\citet{Gua16}} are summarised in figure \ref{fig:alpha_discfracs}. {We expect disc fractions at lower $F_\mathrm{FUV}$ are overestimated as we do not consider other disc dispersal processes (such as internal photoevaporation, see Section} \ref{sec:df_final}). {We therefore focus on matching disc fractions in regions of higher $F_\mathrm{FUV}$. We find that $\tau_\mathrm{gas}=2.5$~Myr and $\alpha= 10^{-2}$ give a good fit to the data for projected $F_\mathrm{FUV} \gtrsim 3000 \, G_0$. Since we have assumed $100 \%$ extinction of FUV photons before gas is completely expelled, this $\alpha$ is an upper limit. As discussed above, reducing the expulsion timescale to $\tau_\mathrm{gas}= 1$~Myr while decreasing the viscosity such that $\alpha = 5 \times 10^{-3}$ also yields the correct disc fractions. However, if gas ejection was initiated at the time of formation of the most massive stars ($\sim 2$~Myr) this would suggest instantaneous expulsion, which is not physical. In this case early supernovae may be responsible for driving gas mass loss (see discussion in Section} \ref{sec:stellpop}). However, for short $\tau_\mathrm{gas}$ Cyg OB2 must have had a extremely large initial stellar mass to maintain the present day central density  ($\gg 10^5 \, M_\odot$, see Section \ref{sec:centmass} and figure \ref{fig:cmass}), which is neither supported by observations nor computationally practicable for the range of models we wish to explore.

While the simplified model presented in this section is not an accurate representation of the dynamical conditions in the region, it represents a first-order approximation on which we can base our choice of kinematic parameters in Section \ref{sec:kinresults}. We will again consider the disc fractions for a more realistic model in Section \ref{sec:df_final}. 

\subsection{Stellar population properties}
\label{sec:kinresults}

{In this section we consider the initial conditions for our $N$-body models of the stellar population of Cyg OB2. We proceed by first aiming to reproduce the observed velocity dispersion in the region by varying the initial velocity distribution  and substructure} (Section \ref{sec:vdisp_res}). Subsequently we match the observed central density by varying the initial stellar mass (Section \ref{sec:centmass}).

\subsubsection{Velocity dispersion, anisotropy and substructure}
\label{sec:vdisp_res}

\begin{figure}
       \includegraphics[width=0.5\textwidth]{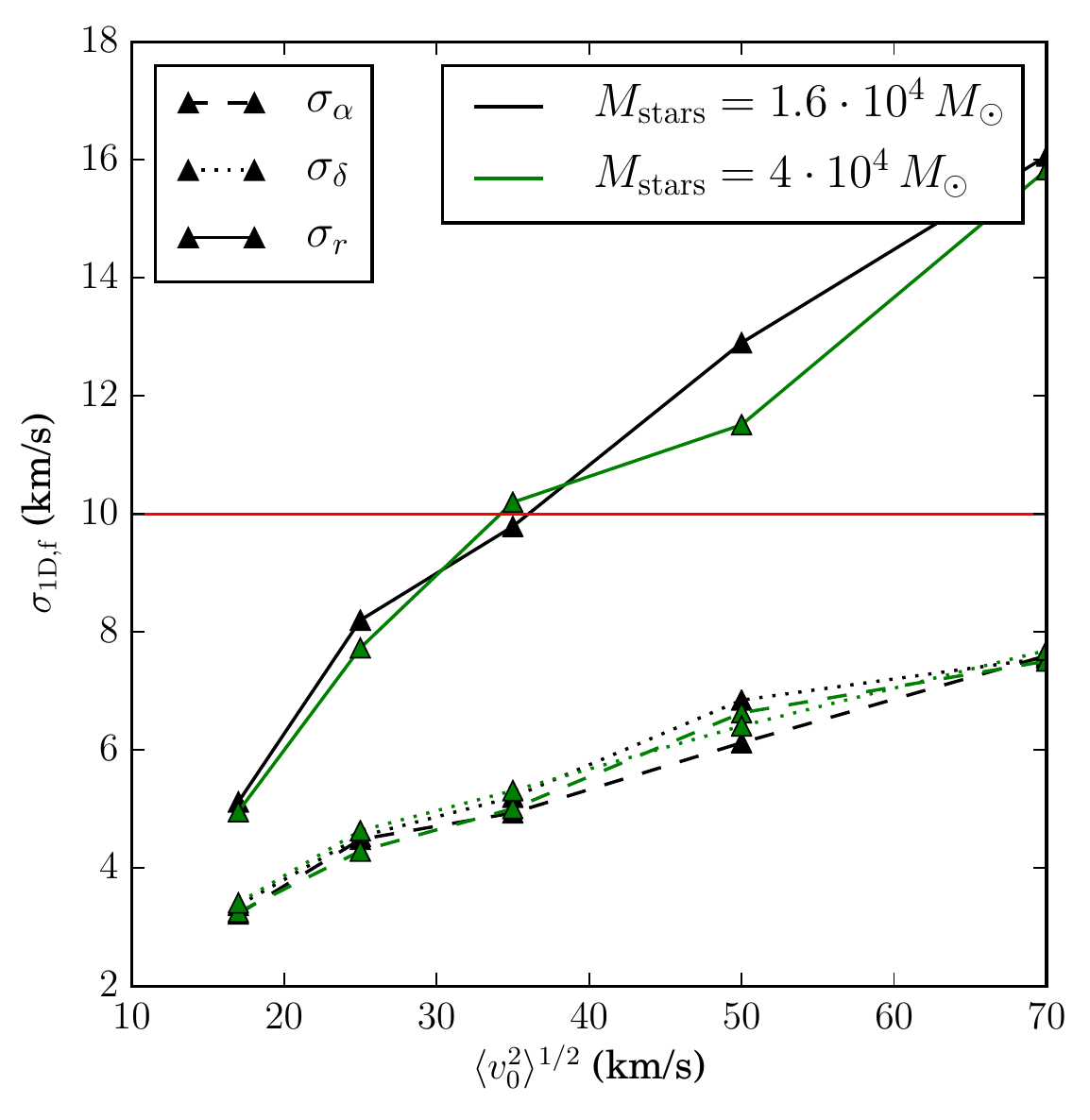}
     \caption{The components of the velocity dispersion ($\sigma_\alpha$, $\sigma_\delta$ and $\sigma_r$, where $\sigma_r$ is the line of sight component) in our model using the \citet{Wri16} field of view after $3 $~Myr of evolution versus the initial three dimensional velocity dispersion over the entire cluster, $\langle v_0^2 \rangle ^{1/2}$. The initial conditions are non-substructured (\texttt{UNIF}) and have initial parameters $a_\mathrm{gas}=10$~pc, $a_\mathrm{stars}=7$~pc, $\tau_\mathrm{exp} = 1.5$~Myr and two different stellar masses $M_\mathrm{stars} = 1.6 \cdot 10^4 \, M_\odot$, $4 \cdot 10^4 \, M_\odot$ (black and green lines respectively). The horizontal red line represents the mean observed 1D velocity dispersion $\langle v^2 \rangle^{1/2}/\sqrt{3} \approx 10$~km~s$^{-1}$. The radial (line of sight) velocity dispersion $\sigma_r > \sigma_{\alpha,\delta}$ due to projection effects and velocity sorting. {Observationally we require a model such that the 1D velocity dispersion components are of the same order (in fact observations indicate $\sigma_{\alpha, \delta}\gtrsim \sigma_r$). This is not reproduced by the \texttt{UNIF} model.} }
     \label{fig:vdispUNIF}
   \end{figure}
   
   \begin{figure}
       \includegraphics[width=0.5\textwidth]{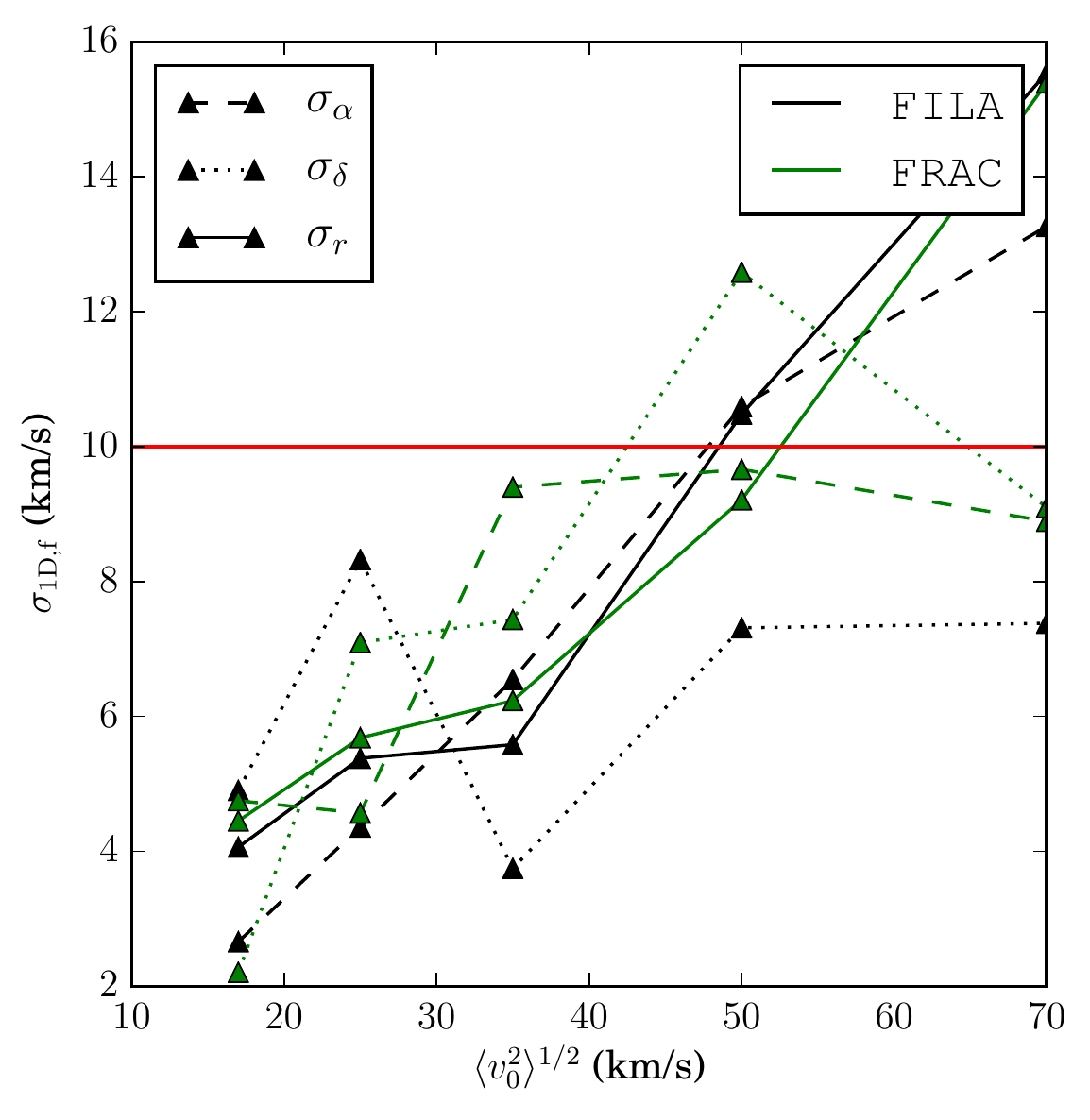}
     \caption{The components of the velocity dispersion  ($\sigma_{\alpha, \delta, r}$) as in figure \ref{fig:vdispUNIF} but for a fixed stellar mass $M_\mathrm{stars} = 1.6 \cdot 10^4 \, M_\odot$. Instead of the \texttt{UNIF} model in figure \ref{fig:vdispUNIF}, the black lines represent the case for a \texttt{FILA} cluster model and the green lines are for a \texttt{FRAC} cluster model, both with $P_0=1$ and $D_0=2.5$. {In this case the large scale substructure gives rise to much greater stochastic variations in the relative 1D velocity dispersions, and is consistent with observations.}}
     \label{fig:vdispFF}
   \end{figure}
   
      \begin{figure}
       \includegraphics[width=0.5\textwidth]{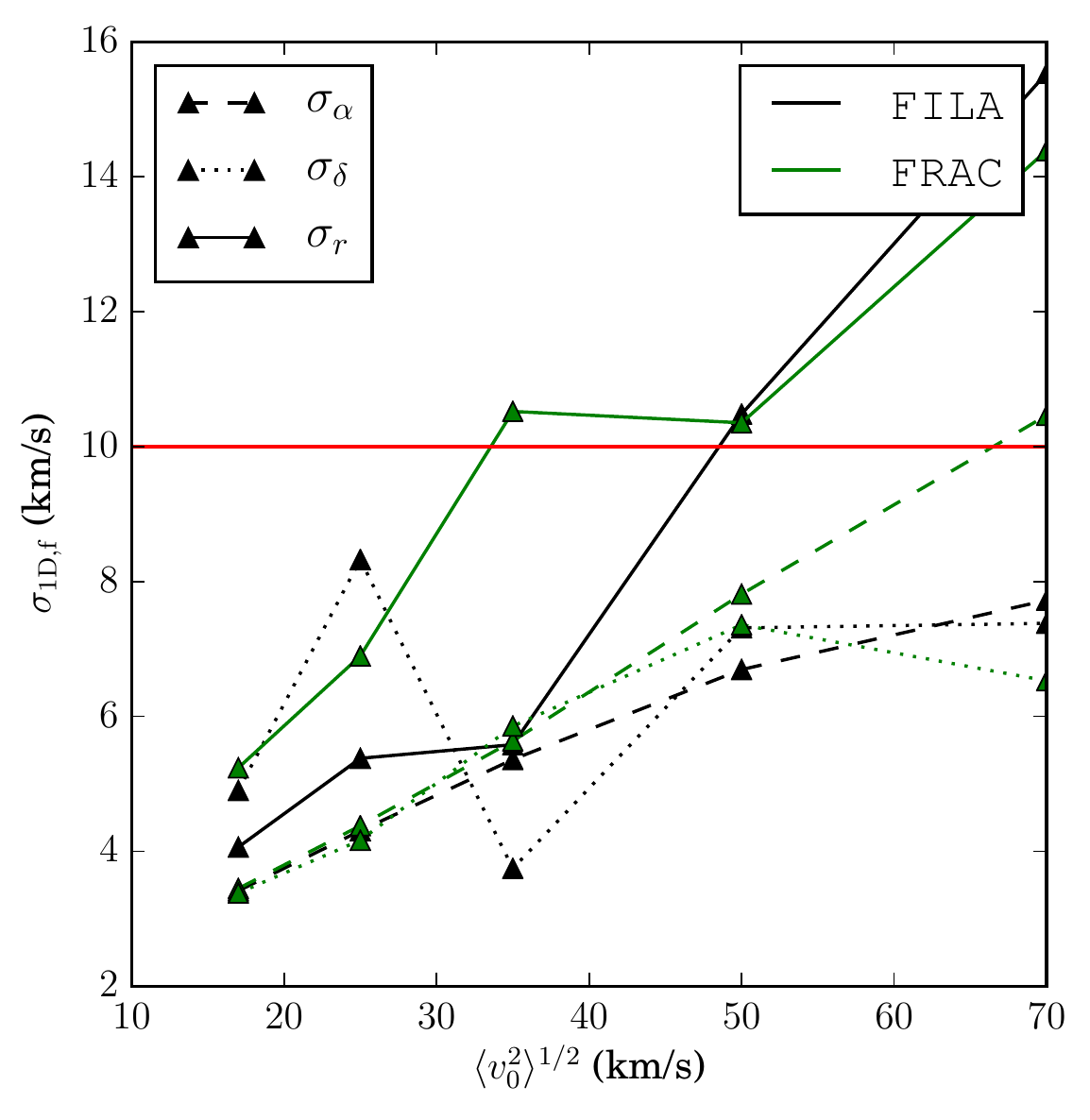}
     \caption{The same as in figure \ref{fig:vdispFF} but with smaller scale substructure ($P_0 = 2$). {In this case the scale of substructure is  insufficient to result in proper motion velocity dispersions comparable to the radial dispersion (we have $\sigma_{\alpha, \delta} < \sigma_r$).}}
     \label{fig:vdispFFP02}
   \end{figure}
   
{In terms of the stellar population, the first observable quantity we aim to reproduce is the central velocity dispersion and its anisotropy} \citep[$\langle v^2 \rangle ^{1/2} \approx 17$~km/s, and the proper motion dispersions $\sigma_{\alpha, \delta}\gtrsim \sigma_r$ the line of sight component -- see][]{Wri16}. This is because it is not strongly sensitive to the initial stellar mass of the cluster. We demonstrate this by considering \texttt{UNIF} cluster models (where cluster conditions are relatively non-stochastic) for a range of initial 3D velocity dispersions and stellar masses; we plot the resuts in figure \ref{fig:vdispUNIF}. As the potential is dominated by the gas in the cluster, and the stellar component is itself highly supervirial, the final central velocity dispersion is insensitive the total stellar mass. 

Figure \ref{fig:vdispUNIF} demonstrates that \texttt{UNIF} models fail to reproduce the observed velocity dispersion. This is because in a non-substructured model we find supressed proper motion velocity dispersions $\sigma_{\alpha,\delta}$ with respect to the radial component $\sigma_r$, since stars with high velocities in the plane of the sky preferentially leave the central field of view over time (velocity sorting). To remedy this, we need to incorporate substructure into our model. 


To demonstrate the influence of substructure on the components of the velocity dispersion within a finite field of view, we run \texttt{FILA} and \texttt{FRAC} models with large scale substructure ($P_0=1$), and rotate a snapshot at $3$~Myr such that the radial velocity dispersion is approximately minimised. The results of this process are shown in figure \ref{fig:vdispFF}, {where the decomposed velocity components are again compared with the initial three dimensional velocity dispersion.} The anisotropies in the velocity dispersion are reproduced in both \texttt{FILA} and \texttt{FRAC} cluster models. 

The degree of anisotropy in the velocity dispersion is dependent on the scale of substructure. Figure \ref{fig:vdispFFP02} shows the results for initially smaller filaments/clumps (using $P_0=2$). While there still exist stochastic fluctuations in the components of the velocity dispersion, the degree of anisotropy is not sufficient to yield proper motion velocity dispersions greater than the radial component. This suggests that the {initial region consisted of large clumps or filaments of mass} $\sim 10^4$~$M_\odot$.

We estimate that the required initial velocity dispersion in Cyg OB2 is $\langle v_0^2 \rangle^{1/2} \sim 50$~km/s, although the stochasticity of the substructured models makes precise estimates impractical. In future models we will assume this is the initial three dimensional velocity dispersion. Other required parameters, such as the initial stellar mass, are chosen to be consistent with this property.

\subsubsection{Central mass}
\label{sec:centmass}
\begin{figure*}
     \subfloat[\label{subfig:cmass2e4} $M_\mathrm{stars} = 2 \cdot 10^4 \, M_\odot$] { \includegraphics[width=0.5\textwidth]{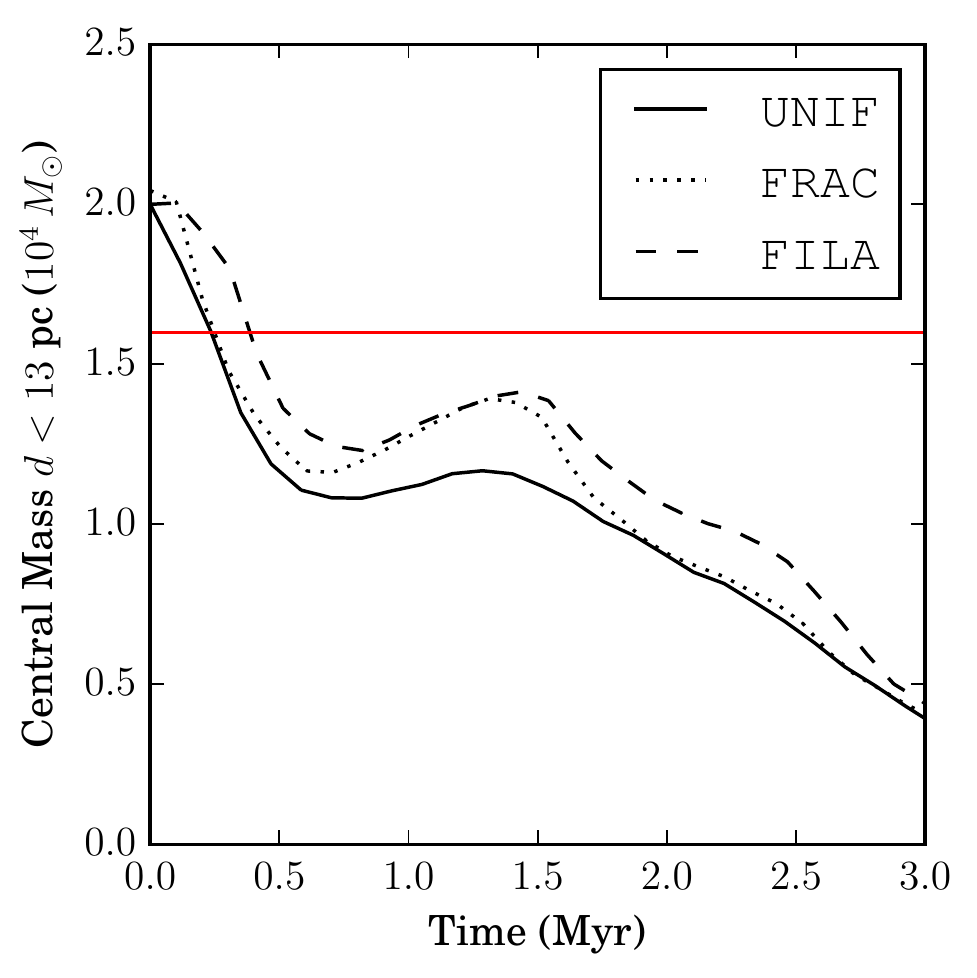}
     }
     \subfloat[\label{subfig:cmass4e4} $M_\mathrm{stars} = 4 \cdot 10^4 \, M_\odot$] { \includegraphics[width=0.5\textwidth]{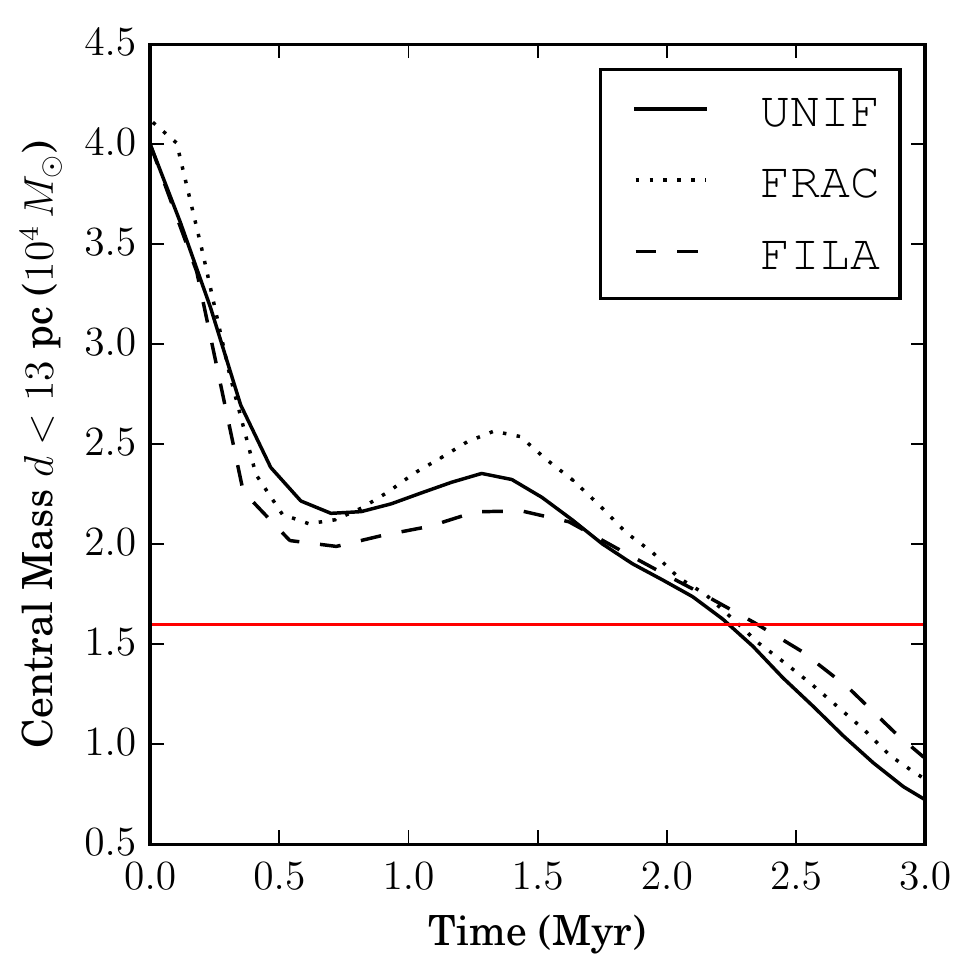}
     }
     
     \subfloat[\label{subfig:cmass8e4} $M_\mathrm{stars} = 8 \cdot 10^4 \, M_\odot$] { \includegraphics[width=0.5\textwidth]{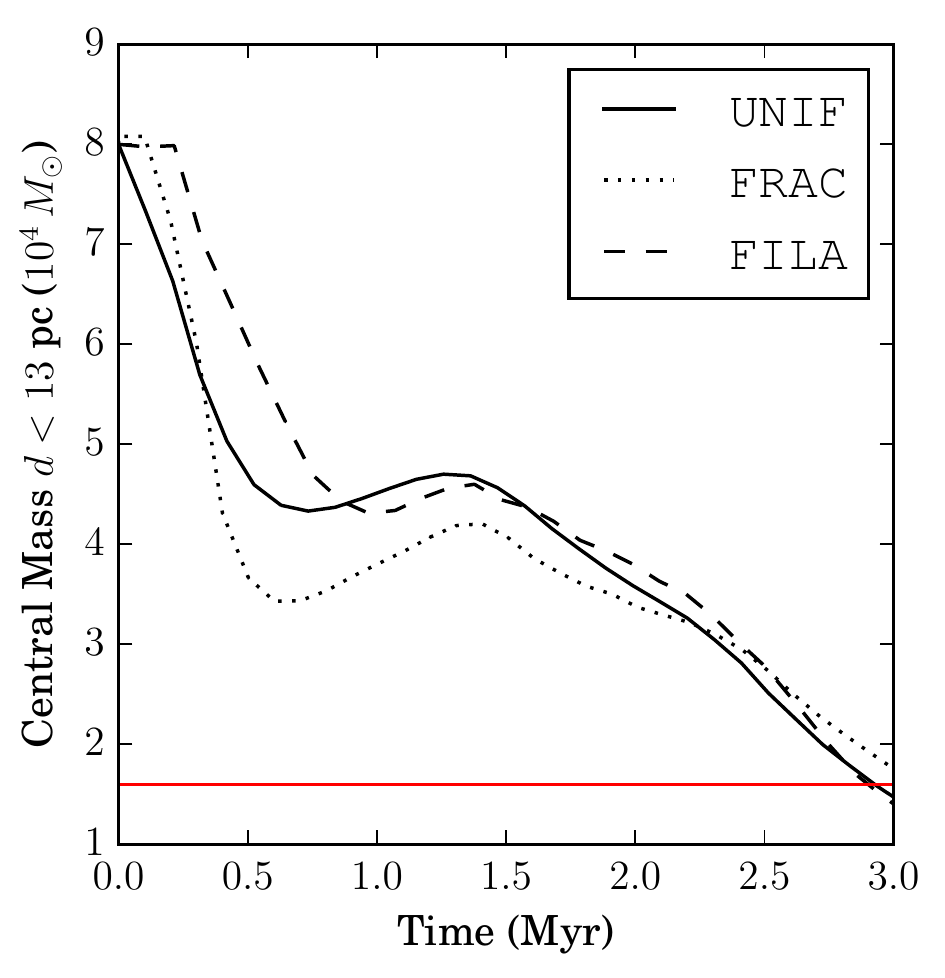}
     }
     \caption{Evolution of the stellar mass within a projected distance of $13$~pc from the centre of the cluster. We show the results for \texttt{UNIF}, \texttt{FRAC} and \texttt{FILA} cluster models (solid, dotted and dashed respectively) over $3$~Myr. All models have initial parameters $a_\mathrm{gas}=10$~pc, $a_\mathrm{stars}=7$~pc,  $\tau_\mathrm{delay} = 1.5$~Myr, $\tau_\mathrm{exp} = 1.5$~Myr and varying stellar mass. The horizontal red line indicates the observed central mass $\sim 1.6 \times 10^4 \, M_\odot$ \citep{Wri15}. {We find that an initial mass of $\sim 8 \times 10^4 \, M_\odot$ reproduces the observed central density.}}
     \label{fig:cmass}
   \end{figure*} 
   
{We wish to alter the initial stellar mass such that the central density at the end of the simulation is consistent with observations.} A cluster model with a central mass of $1.6 \times 10^4 \, M_\odot$ after $3$~Myr of evolution is required, where $\langle v_0^2 \rangle^{1/2} = 50$~km/s as discussed in Section \ref{sec:vdisp_res}. To find the appropriate initial mass we run \texttt{UNIF}, \texttt{FRAC} and \texttt{FILA} models at stellar masses $M_\mathrm{stars} = 2 \times 10^4 \, M_\odot$, $4 \times 10^4 \, M_\odot$ and $8 \times 10^4 \, M_\odot$. The results are shown in figure \ref{fig:cmass}. We find that $M_\mathrm{stars} = 8 \times 10^4 \, M_\odot$ (figure \ref{subfig:cmass8e4}) is sufficient to yield the required central mass after $3$~Myr of evolution. 

Neither the initial mass nor the substructure has a significant effect on the fraction of the stellar mass which remains within a $13$~pc projected radius (approximately a quarter in each case). As in the case of our velocity investigation, this is due to the fact that that the stellar component of the potential energy is much smaller than the total kinetic energy. 

{The central mass in all of our models undergoes a similar temporal evolution. In figure} \ref{fig:cmass} {we see an initial rapid mass loss as stars with the highest energies escape the potential. This is because we do not truncate the velocity dispersion such that escapers are initially forbidden; it is not clear whether or not this is physically realistic, and the effect of choosing such initial conditions is discussed in Section} \ref{sec:summary_bfm} in the context of primordial gas mass. However, since stars which escape the central regions are not considered in our PPD models and the gravitational potential is dominated by the gas component, whether or not these early escapers are initially included in the model is of secondary importance. {After this initial decline in mass, some high energy stars remain bound, and therefore return to the central regions, causing a modest oscillation in the mass. This is again a consequence of using a Boltzmann velocity distribution without truncating the high velocity end. The magnitude of this oscillation is more variable for substructured regions where velocities (and therefore kinetic energies) are correlated. Gas expulsion (starting at $\tau_\mathrm{delay} = 1$~Myr and continuing over $\tau_\mathrm{exp}= 1.5$~Myr) results in a decrease in gravitational potential, and as the number of stars with energies sufficient to escape increases, the central mass decreases. }

\subsubsection{Summary of best fitting model}
\label{sec:summary_bfm}
   
\begin{figure*}
     \subfloat[\label{subfig:vfield_clump}] { \includegraphics[width=0.5\textwidth]{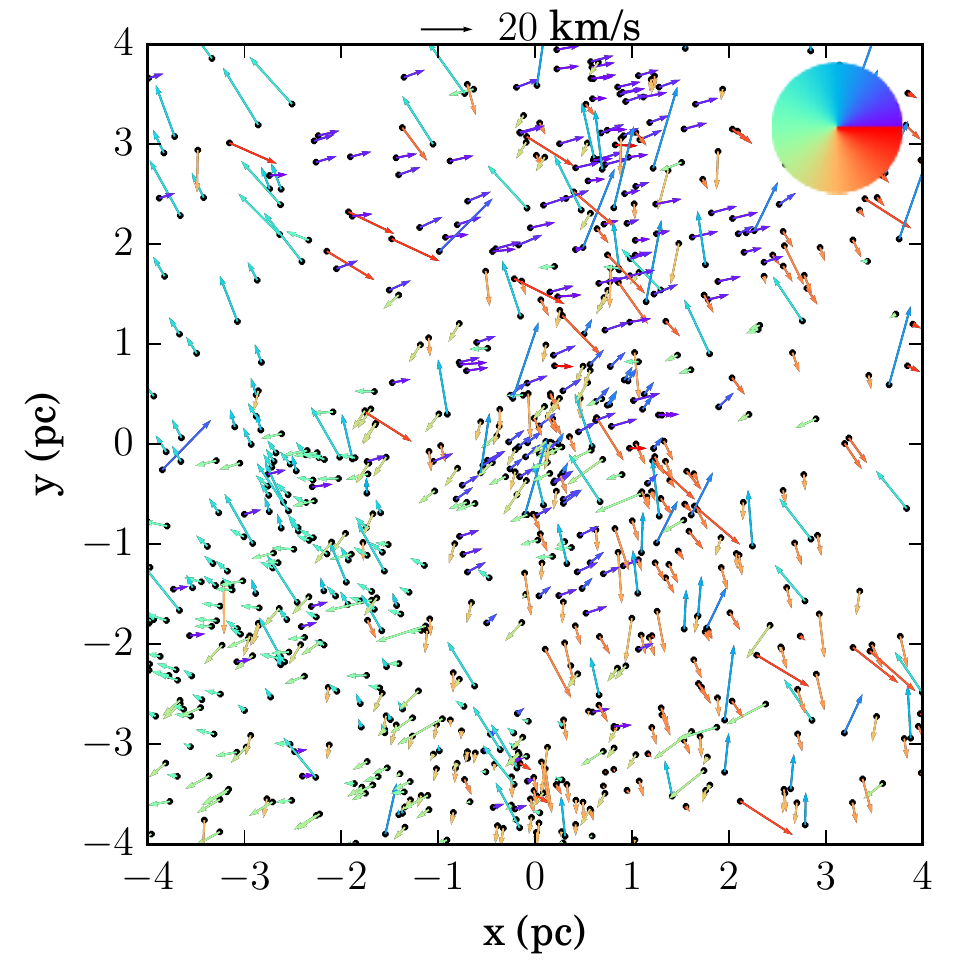}
     }
     \subfloat[\label{subfig:vfield_rad}] { \includegraphics[width=0.5\textwidth]{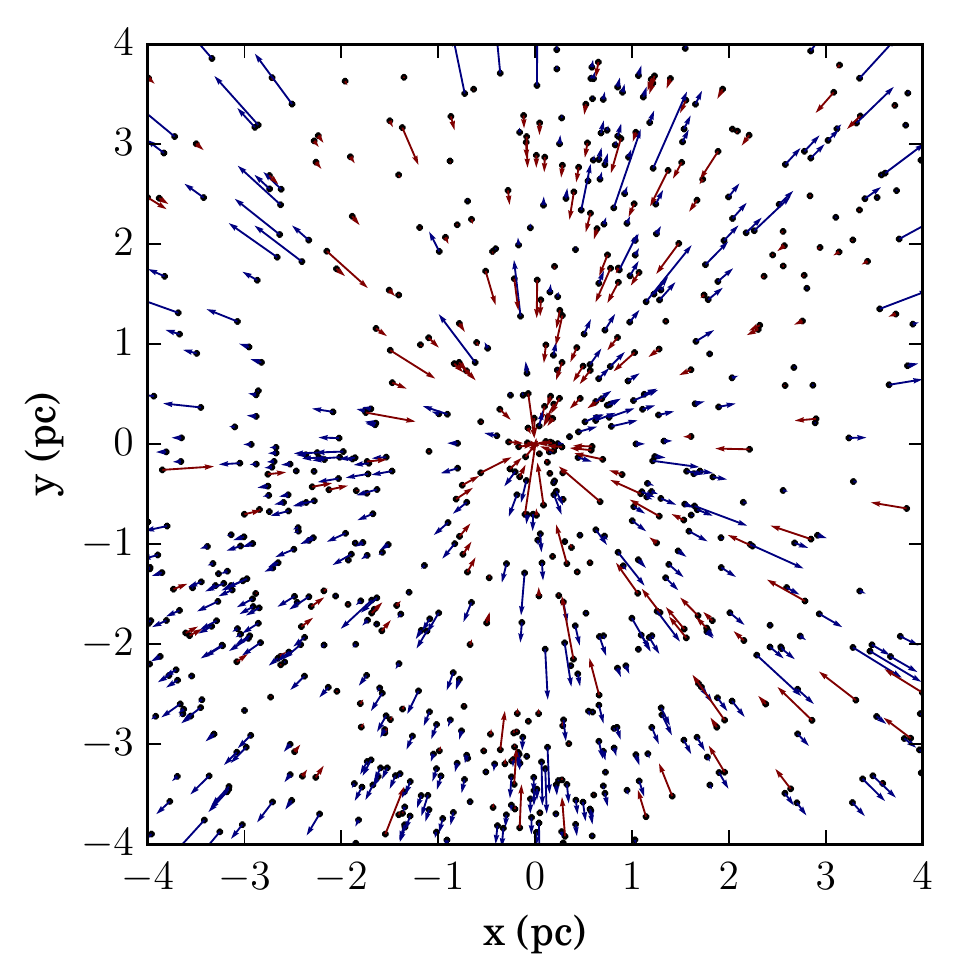}
     }
     \caption{Velocity field of a subset of $800$ stars in the central region of our chosen model (summarised by the parameters in Table \ref{table:cygparams}) after $3$~Myr of evolution. In figure \ref{subfig:vfield_clump} velocity vectors are colour coded by their direction to illustrate the underlying substructure; a correlation can be seen between position and velocity vectors. In figure \ref{subfig:vfield_rad} only the radial components in the plane of the sky are shown, coloured blue for stars moving outwards from the centre and red for infalling stars. There is no clear bias between infalling and outgoing velocities - see text for details. Similarly, \citet{Wri16} {found correlations between position and velocity vectors, and that Cyg OB2 shows no sign of expansion from the apparent centre.} }
     \label{fig:vfield}
   \end{figure*}  

\begin{table*}
\centering 
 \begin{tabular}{c c c c c c c c c c } 
 \hline
Type & $M_\mathrm{stars}$ ($M_\odot$) &  $a_\mathrm{stars}$ (pc) & $\gamma$ & $P_0$ & $D_0$ & $M_{\mathrm{gas, }0}$ ($M_\odot$) & $a_\mathrm{gas}$ (pc) & $\tau_\mathrm{delay}$ (Myr) & $\tau_\mathrm{exp}$ (Myr)  \\ [0.5ex] 
 \hline
 \texttt{FILA} & $8 \cdot 10^4$ &  $7$ & $5.8$ & $1$ & $2.5$ & $7.9 \cdot 10^6$ & $10$ & $1$ & $1.5$   \\
 [1ex] 
 \hline
\end{tabular}
\caption{Parameters of the `best-fit' model, used to reproduce the properties of the observed stellar population of Cyg OB2.} 
\label{table:cygparams}
\end{table*}

By considering gas expulsion, stellar mass, velocity dispersion and initial substructure, we have found a model for the evolution of Cyg OB2, summarised by the parameters in Table \ref{table:cygparams}, which fits observations of the stellar population. {While we refer to this as our `best-fit' model, this is to be understood as the result our process of deduction in terms of the appropriate parameters, and not as the optimisation of a statistical metric or parameter space exploration.} In our models we find gas expulsion was completed $\sim 0.5$~Myr ago (for a PPD viscous timescale of $0.5$~Myr), the initial velocity dispersion was $\sim 50$~km/s and the initial cluster mass was $\sim 8 \times 10^4 \, M_\odot$. We further suggest that the largest scales of initial coherent clumps within the primordial Cyg OB2 had a mass of $\sim 10^4 \, M_\odot$. No significant dynamical differences can be found between \texttt{FRAC} and \texttt{FILA} type models at the present time, and we hereafter use a \texttt{FILA} model in our analysis of the disc population (Section \ref{sec:df_final}).

The gas mass required for initial virial equilibrium is $M_\mathrm{gas} \sim 8 \cdot 10^6 \, M_\odot$ (with a scale parameter $a_\mathrm{gas} = 10$~pc). This would make the primordial GMC massive compared to known Milky Way molecular clouds, although the census is not complete \citep[see][for a review]{Lon14}. It is also possible that the initial velocity dispersion is overestimated due to the truncation of equation \ref{eq:MBdist} at high velocities \citep[simulations suggest stars have a subvirial initial velocity dispersion with respect to the primordial gas - see][]{Off09,Kru12}. While this would not influence the central velocity dispersion (high velocity stars leave the centre in any case) it would reduce the number of escapers early on in the cluster evolution and therefore the required initial stellar mass. It would also reduce the gas mass necessary for virial equilibrium as $M_\mathrm{gas} \propto \langle v^2 \rangle$ if $M_\mathrm{gas} \gg M_\mathrm{stars}$ (equation \ref{eq:totalpot}). If our gas mass estimate is accurate then this makes the star formation efficiency $\sim 1\%$, although this is probably a lower limit (an upper limit on $M_\mathrm{gas}$).

\subsubsection{Substructure and expansion observables}
\label{sec:bfmod_expsub}
Alternate kinematic constraints not considered in the previous analysis include the measures of substructure and the absence of expansion signatures in the stellar kinematics \citep{Wri16}. We find that determining these metrics is problematic for a given cluster model. This is because the values obtained differ stochastically depending on initial conditions, the time of `observation' and the subset of stars used in taking a measurement. To obtain an accurate probability of finding the observed values for these metrics, a large number of models would need to be tested, which would be computationally expensive. However, for our chosen model the velocity field is illustrated in figure \ref{fig:vfield}. We find that the correlation between positions and proper motions is clear (figure \ref{subfig:vfield_clump}), while figure \ref{subfig:vfield_rad} does not show clear evidence of expansion. However, the stellar components are in fact expanding globally since they are unbound. 

  \begin{figure}
       \includegraphics[width=0.5\textwidth]{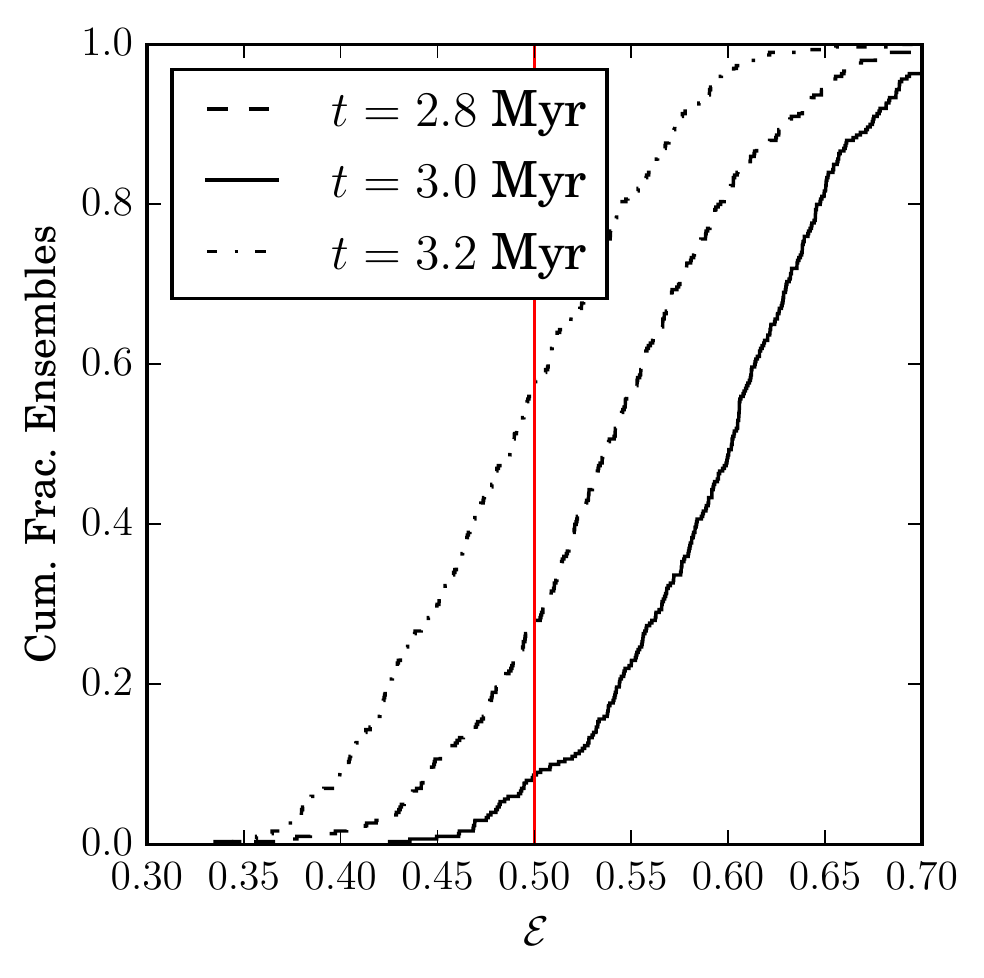}
     \caption{The cumulative fraction of the number of ensembles (each defined to be a subset of $800$ stars in the \citealt{Wri16} field of view) with expansion parameter $\mathcal{E}$ (equation \ref{eq:expansion}) in our chosen model. A value of $\mathcal{E}\approx 0.5$, which was found by \citet{Wri16} for Cyg OB2, would observationally be taken as an indication that no expansion is occuring. }
     \label{fig:expansioncumfrac}
   \end{figure}

{To illustrate this point more fully, we define the expansion parameter:}
\begin{equation}
\label{eq:expansion}
\mathcal{E} = \frac{T^+}{T^- + T^+}
\end{equation} {where $T^{+/-}$ is defined as the total kinetic energy of stars directed in the postive/negative projected radial direction (in the plane of the sky). Thus $\mathcal{E} \rightarrow 1$ or $0$ if the velocity dispersion indicates rapid expansion or contraction respectively. A value of $\mathcal{E}\approx 0.5$ would usually be taken as evidence that a stellar population is not expanding. In Figure} \ref{fig:expansioncumfrac} {we show the cumulative distribution of the measured expansion parameter for random subsets of $800$ stars in the} \citet{Wri16} { field of view (the central $8$~pc~$\times$~$8$~pc). {We find that a wide range of values for $\mathcal{E}$ is possible at any given time. Depending on the time at which the velocities are observed and the chosen subsample, our model is found to be consistent with an observed value $\mathcal{E}\approx 0.5$}. This is because the $\mathcal{E}$ distribution varies considerably and non-monotonically in time even for a single model. However, at any given time the stellar population is expanding (filaments are moving away from each other). Alternative geometrical signatures may be more successful at gauging such expansion. Due to the stochasticity and wide range of possible $\mathcal{E}$ values for our model, we conclude that $\mathcal{E}$ alone is not a sufficient metric to draw conclusions on the expansion of a substructured association.} For further discussion on the kinematic indicators of expansion in OB associations, see \citet{Bau07} and \citet{War18}.   

In the remainder of this work we will first revisit the disc population in our model, checking that the population is consistent with the known disc fractions in the region. Subsequently we will explore predictions for the disc mass and radius distribution relevant for future observations in the region. 

\subsection{Disc properties}
\label{sec:df_final}

In what follows we will consider a \texttt{FILA} model with the properties described in Table \ref{table:cygparams}. We fix the $\alpha$-viscosity of the disc population with the derived value $\alpha = 10^{-2}$ (Section \ref{sec:df_res1}). As in Section \ref{sec:df_res1} all discs are assumed to have an initial mass that is uniformly distributed between $1$ -- $10 \%$ of their host star mass (see Section \ref{sec:visc_evol}), and scale radius as in equation \ref{eq:R1}. We consider the same range of stellar masses $0.5$ -- $2$~$M_\odot$ and calculate the evolution of a subset of $5000$ discs. 

\subsubsection{Disc fractions in best-fit model}
\begin{figure*}
     \subfloat[\label{subfig:g0bar} ] { \includegraphics[width=0.5\textwidth]{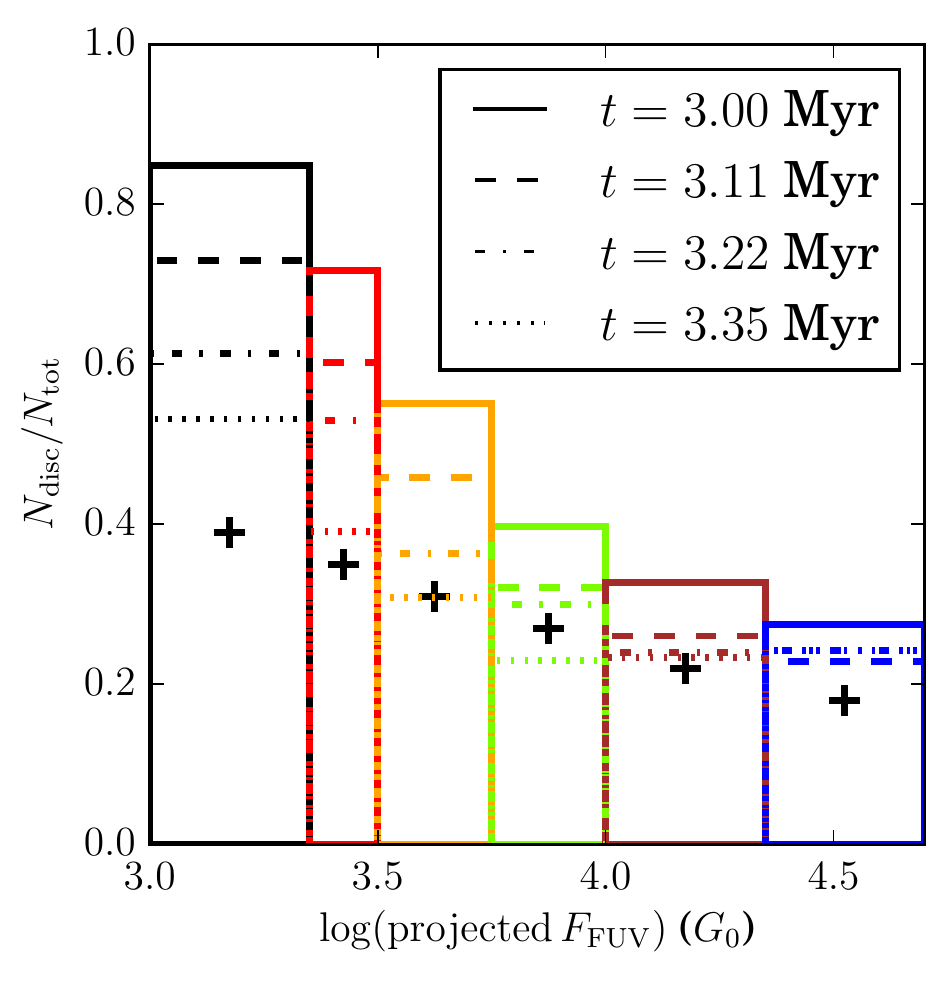}
     }
     \subfloat[\label{subfig:g0phys} ] { \includegraphics[width=0.5\textwidth]{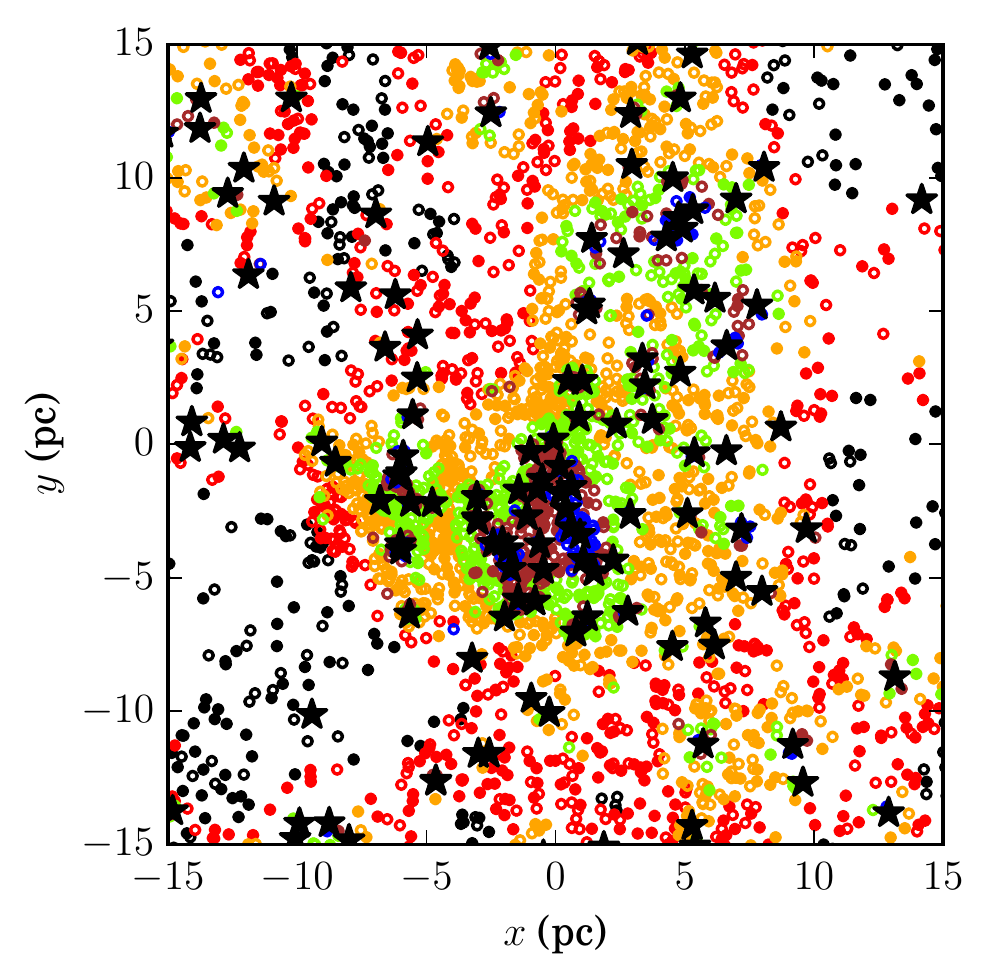}
     }
     \caption{Model of the disc population in a model described by the parameters in Table \ref{table:cygparams}.  In figure \ref{subfig:g0bar} we show the  disc fraction as a function of FUV flux, calculated by projected distance to massive stars, at varying times. These fractions are in good agreement with the observed disc fractions, indicated by black crosses. In figure \ref{subfig:g0phys} we show the physical distribution of the disc population after $3$~Myr, colour coded by the projected FUV flux as in figure \ref{subfig:g0bar}. Star markers represent the positions of stars with a mass $> 10 \, M_\odot$. Empty circles represent a disc with a mass $< 10^{-5} \, M_\odot$, while filled circles are `surviving' discs with a greater mass  \citep[c.f. fig. 3 in ][]{Gua16}.}
     \label{fig:discfrac_model}
   \end{figure*}
   
The disc fraction distribution in our best-fit model after $\sim 3$~Myr is summarised by figure \ref{fig:discfrac_model}, in which we show the surviving disc fraction as a function of projected FUV flux (figure \ref{subfig:g0bar}) with the spatial distribution and projected $F_\mathrm{FUV}$ in figure \ref{subfig:g0phys}. We find that after $3$~Myr the number of surviving discs is slightly overestimated in our model, particularly at the lower FUV fluxes. There are two reasons why we expect this to be the case. {Firstly, extinction in the FUV may decrease before gas is fully removed from the cluster.} If gas is expelled due to the flux from massive stars, then the central, highly irradiated environments would become less dense and allow efficient photoevaporation at earlier times. Additionally, clumpy gas distributions may have a similar effect in reducing extinction during this period. Secondly, internal photoevaporation due to the stellar host depletes the gas content even when a PPD evolves in isolation. This speeds up destruction timescales, particularly in regions of lower $F_\mathrm{FUV}$, where the mass loss rates induced by internal and external photoevaporation become comparable.

The influence of the above considerations is uncertain and we therefore do not attempt to model them here. Our model does however reproduce the correct disc fractions within a reasonable period of time, particularly at $F_\mathrm{FUV} \gtrsim 3000 \, G_0$. We conclude that the observations of \citet{Gua16} can be explained by external photoevaporation of the PPD population in Cyg OB2.

\subsubsection{Disc mass and FUV flux environment}
\label{sec:md_res}
\begin{figure*}
     \subfloat[\label{subfig:mg0} ] { \includegraphics[width=0.5\textwidth]{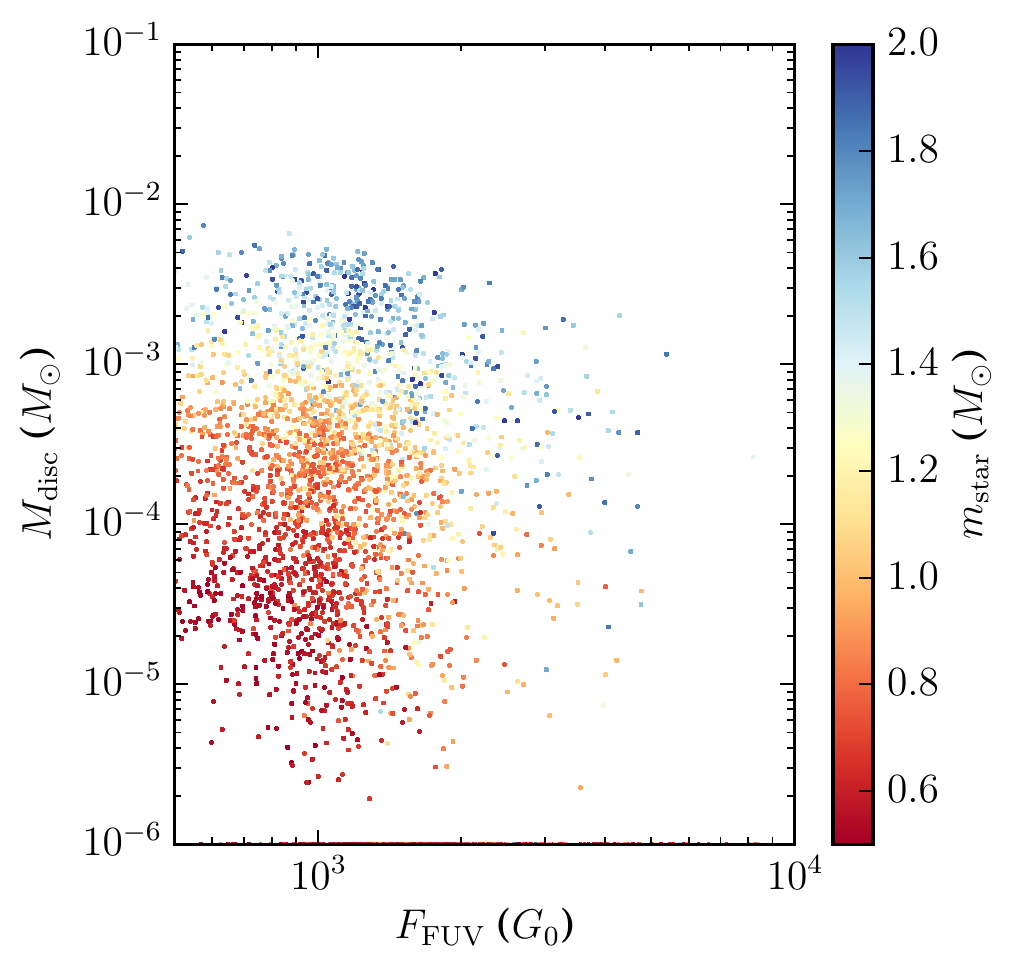}
     }
     \subfloat[\label{subfig:mg0proj} ] { \includegraphics[width=0.5\textwidth]{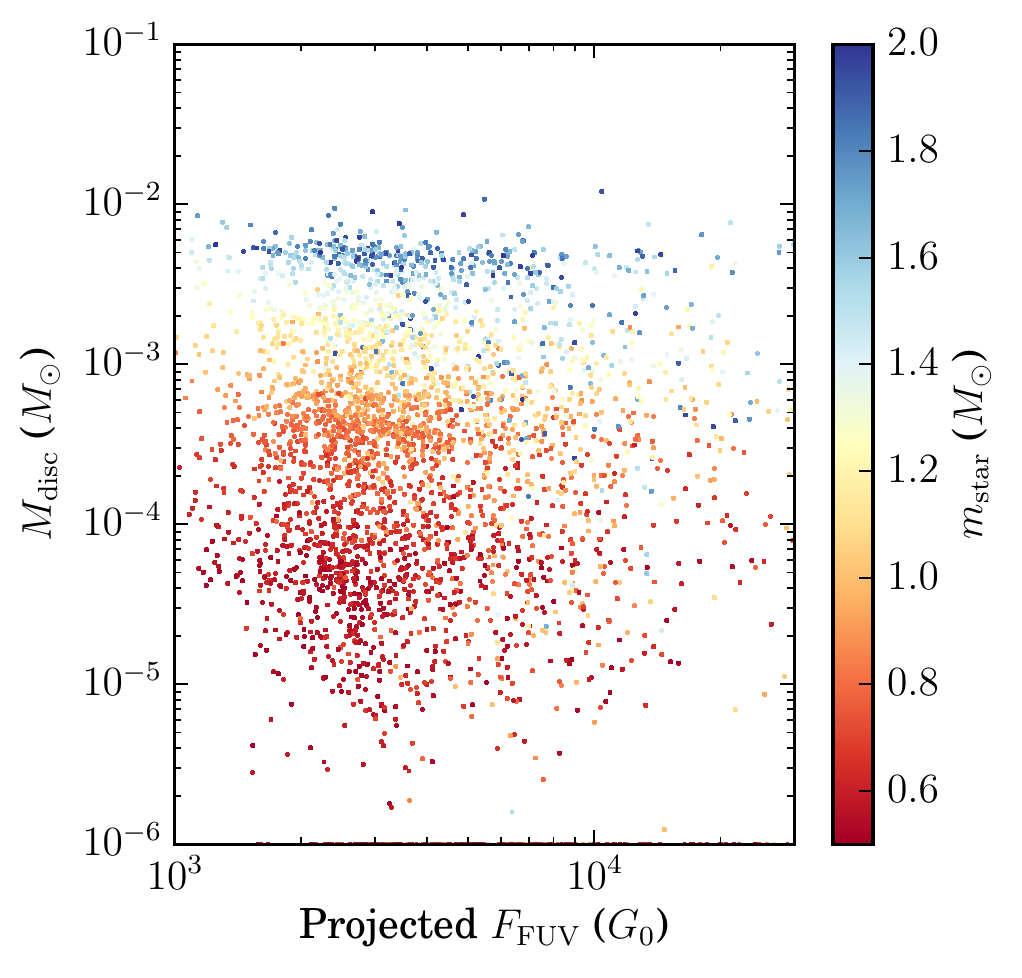}
     }
     \caption{PPD mass after $3$~Myr in our chosen model (described by the parameters in Table \ref{table:cygparams}) as a function of real and projected FUV flux (figures \ref{subfig:mg0} and \ref{subfig:mg0proj} respectively). Points are colour coded by the mass of the host star.  Initial disc masses are drawn from a uniform distribution between $1\%$ and $10\%$ of the host mass. In the context of figure \ref{fig:discfrac_model}, discs with masses $<10^{-5} \, M_\odot$ are considered `destroyed'. {We find that $F_\mathrm{FUV}$ is a poor indicator of disc mass.} }
     \label{fig:mvG0}
   \end{figure*}

   \begin{figure}
   \includegraphics[width=0.5\textwidth]{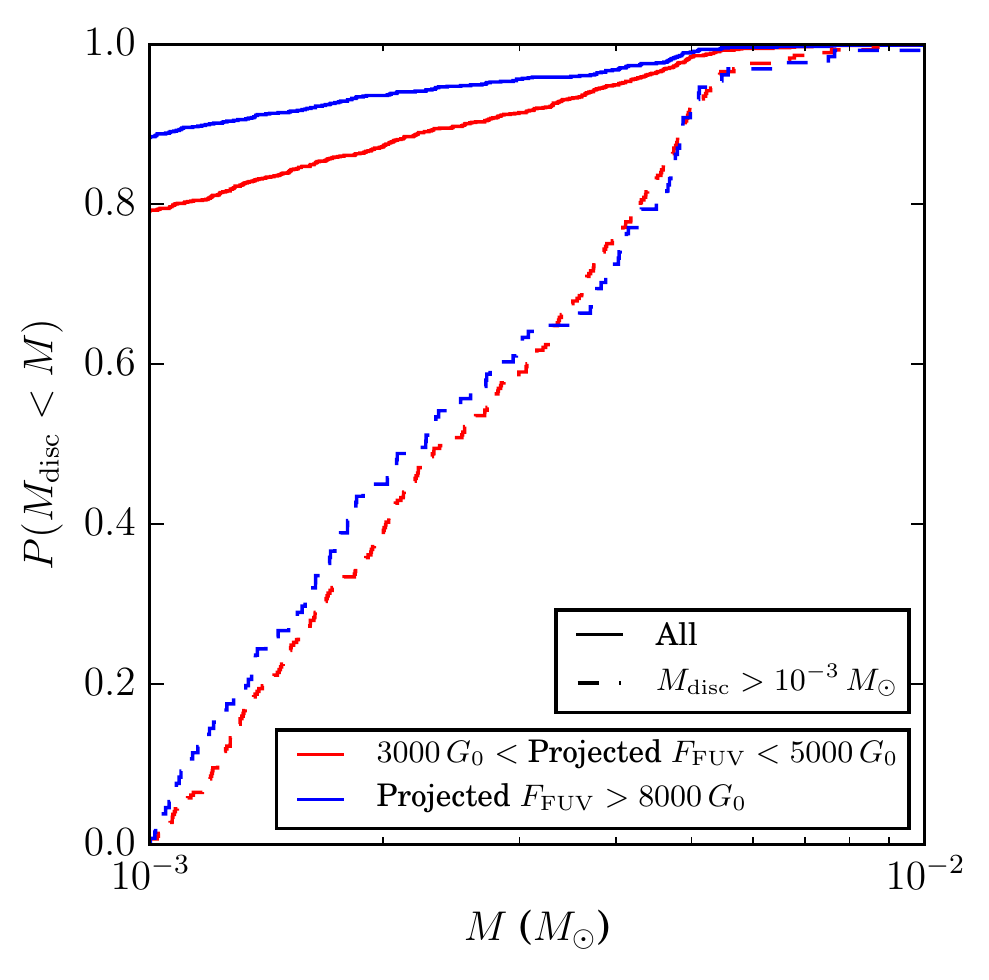}
   \caption{Cumulative fraction of disc mass after $3$~Myr in our chosen model. Solid lines are for the entire disc population, while the dashed lines include only discs which have $M_\mathrm{disc}>10^{-3}\, M_\odot$. The disc population is divided up by projected (observed) FUV flux; the red lines are for stars experiencing an apparent $F_\mathrm{FUV}$ between $3$--$5 \times 10^3 \, G_0$, while the blue lines correspond to stars with  $F_\mathrm{FUV}>8 \times 10^3$. A large sample of all discs would be required to detect differences between the masses in the two FUV flux bins.  }
   \label{fig:cumfracs}
  \end{figure}

To make predictions about the disc population in Cyg OB2 for comparison with future observations we consider PPD mass as a function of FUV flux (figure \ref{fig:mvG0}). We find that, if used in isolation, the local flux experienced by a given star is a poor predictor of the disc mass. Correlation with real or projected FUV flux only becomes clear when the host mass, to which the final PPD mass is closely correlated (as demonstrated by the colour gradient in figure \ref{fig:mvG0} and discussed in Section \ref{sec:mi_res}), is also taken into account. In particular, we expect that low mass stars ($<1\, M_\odot$) in the centre of Cyg OB2 host exclusively low mass discs ($\lesssim 10^{-3} \, M_\odot$, if any).
      
Without taking into account stellar mass, can we find differences in disc properties between PPDs in apparently high versus low FUV flux environments? To answer this, we must consider the sensitivity limit for future observations. Considering \textit{ALMA} band 6 sensitivity, reasonable integration times ($\sim 30$~minutes) for a survey sample suggest flux densities down to $F_\nu (850 \, \mu \mathrm{m})\sim 40$~$\mu$Jy can be detected. At the distance of Cyg OB2 this means that dust masses can be established down to a few $M_{\earth}$ \citep{And05}. The corresponding total disc masses are $M_\mathrm{disc} \sim 10^{-3}\, M_\odot$ if the gas to dust ratio is $\Sigma_\mathrm{gas}/\Sigma_\mathrm{dust}=10^2$. Given that the PPDs are likely to be gas depleted by external photoevaporation, the latter assumption is probably not accurate for many discs \citep{Ans16}, and this complicates the interpretation of observations. Nonetheless, we show the cumulative PPD mass fraction after $3$~Myr for two FUV flux bins in figure \ref{fig:cumfracs}. While disc masses are indeed suppressed at higher projected FUV flux, considering the sensitivity limit of \textit{ALMA} makes finding differences between the two populations impractical. A sample of several $100$s of PPDs would be required to find a difference between the total population in high and low FUV flux environments (including non-detections). We find that similar sample sizes would be required to find differences in disc outer radius distributions in the two environments.

  \subsubsection{Stellar mass independent disc initial conditions}
  \label{sec:mi_res}
  
  \begin{figure}
       \includegraphics[width=0.5\textwidth]{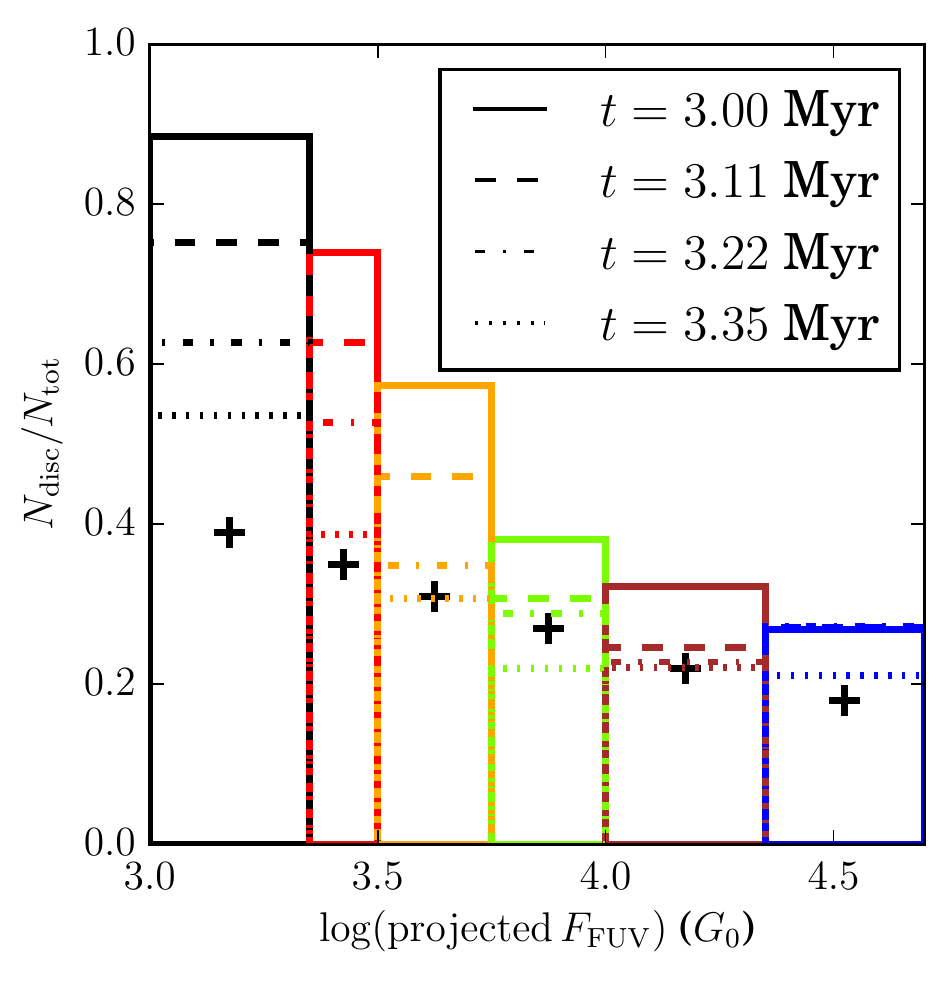}
     \caption{As in figure \ref{subfig:g0bar} but for a distribution of initial disc masses independent of the stellar host mass (see text for details). This demonstrates that reproducing the observed disc fractions is not sensitive to the choice of PPD initial conditions. }
     \label{fig:g0bart_dist}
   \end{figure}
   
  \begin{figure}
     \includegraphics[width=0.5\textwidth]{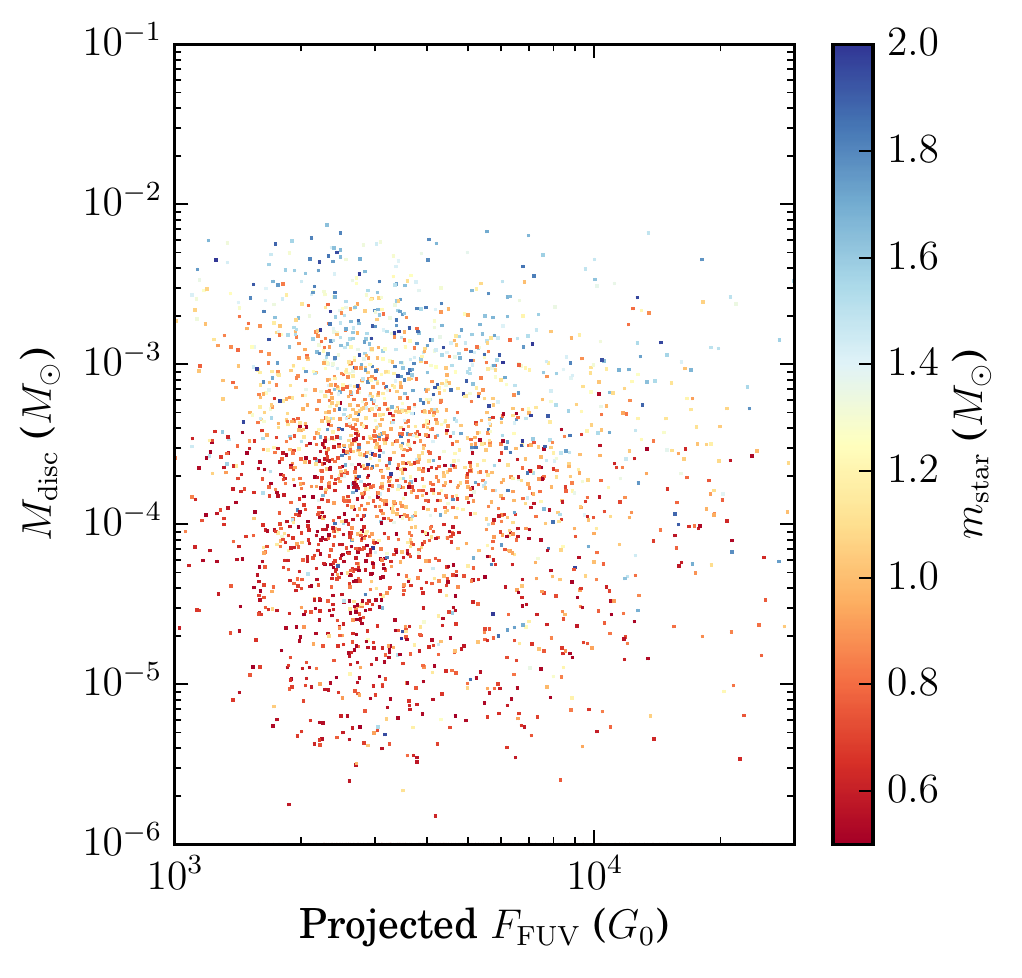}
     \caption{As in figure \ref{subfig:mg0proj} but for a distribution of initial disc masses independent of the stellar host mass (see text for details).}
     \label{fig:mvG0dist}
   \end{figure}

  In obtaining the results in Section \ref{sec:md_res} we have already assumed that the initial disc mass is dependent on the stellar mass of the host. We wish to confirm our results are not sensitive to this assumption. We therefore recalculate the PPD evolution in our model under the assumption that the initial disc mass is uncorrelated to the stellar mass. To this end we draw the initial disc masses from a log-normal distribution:
  $$
  M_{\mathrm{disc},0}/M_{\odot}  =  e^{\mu + \sigma X}
  $$ where the random variable $X \sim \mathcal{N}(0,1)$ is drawn from a standard normal distribution, uncorrelated with $m_\mathrm{star}$. The values of $\mu=-3.25$ and $\sigma=0.7$ are chosen such that the mean and dispersion of $M_{\mathrm{disc},0}$ match those chosen for the stellar mass dependent initial conditions. The scale radius $R_1$ is again defined according to equation \ref{eq:R1}. {This distribution of initial conditions recovers similar disc fractions as a function of projected $F_\mathrm{FUV}$ (figure} \ref{fig:g0bart_dist}). The resulting disc mass distribution is shown in figure \ref{fig:mvG0dist}. Our findings do not differ greatly from those in figure \ref{fig:mvG0}, except for a predictable weaker correlation between final disc mass and host mass. 
  
  \subsubsection{Disc mass dependence on host mass}
  
   \begin{figure*}
     \subfloat[\label{subfig:mstvmdisc} ] { \includegraphics[width=0.5\textwidth]{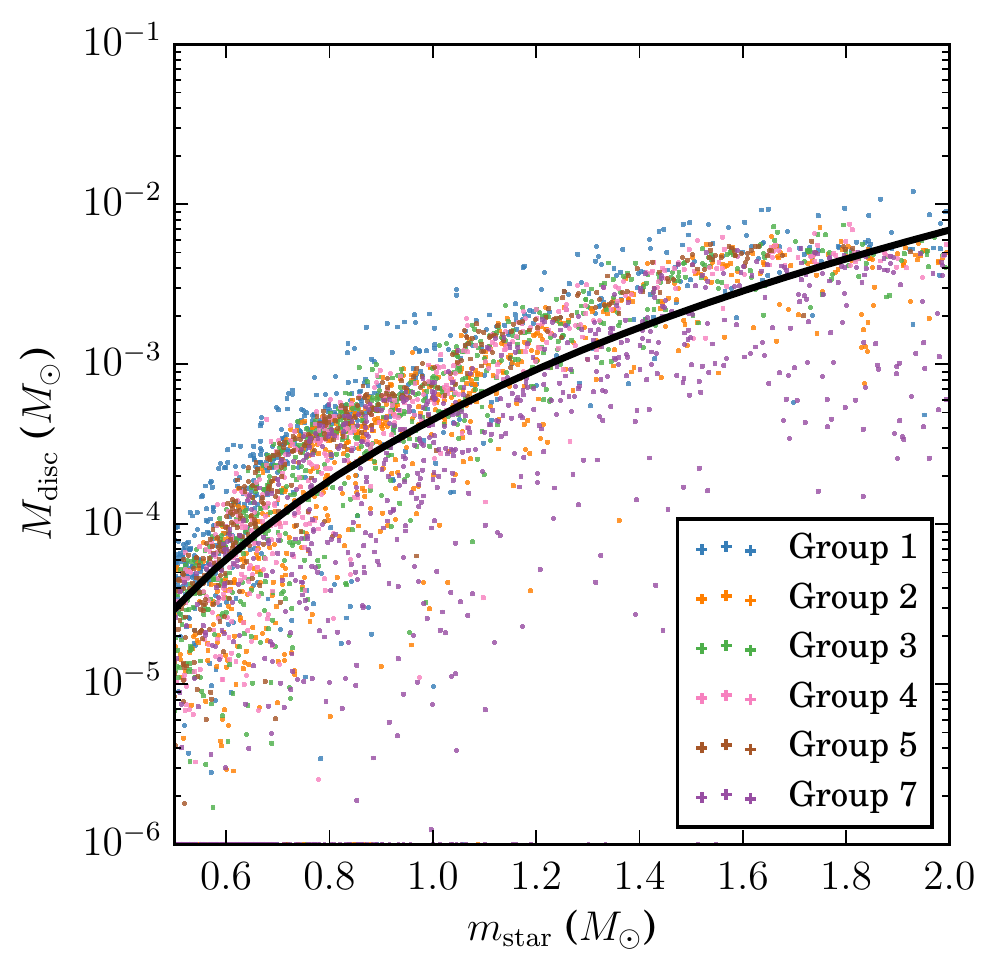}
     }
     \subfloat[\label{subfig:mstvmdisc_dist} ] { \includegraphics[width=0.5\textwidth]{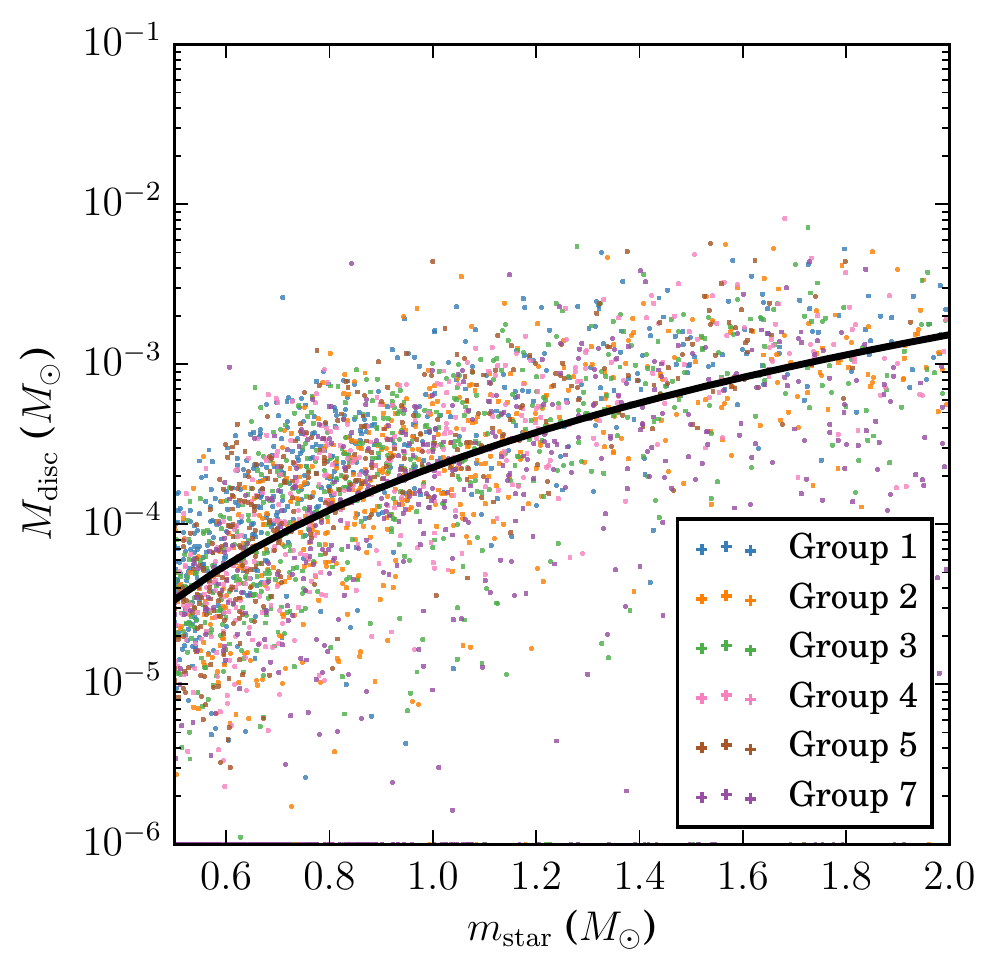}
     }
     \caption{PPD mass distribution after $3$~Myr of evolution in our chosen model (see text and Table \ref{table:cygparams} for details). In figures \ref{subfig:mstvmdisc} and \ref{subfig:mstvmdisc_dist} initial disc masses are dependent on and independent of stellar host mass respectively.  The black line follows $(M_\mathrm{disc}/M_\odot) = 4.5 \cdot 10^{-4} \left(m_\mathrm{star}/M_\odot \right)^{3.9}$ in figure \ref{subfig:mstvmdisc}, and $(M_\mathrm{disc}/M_\odot) = 2.3 \cdot 10^{-4} \left(m_\mathrm{star}/M_\odot \right)^{2.8}$ in figure \ref{subfig:mstvmdisc_dist}. The points are colour coded by the largest scale fractal sub-group to which they belong.}
     \label{fig:mstvdisc}
   \end{figure*}

       \begin{figure}
      \includegraphics[width=0.5\textwidth]{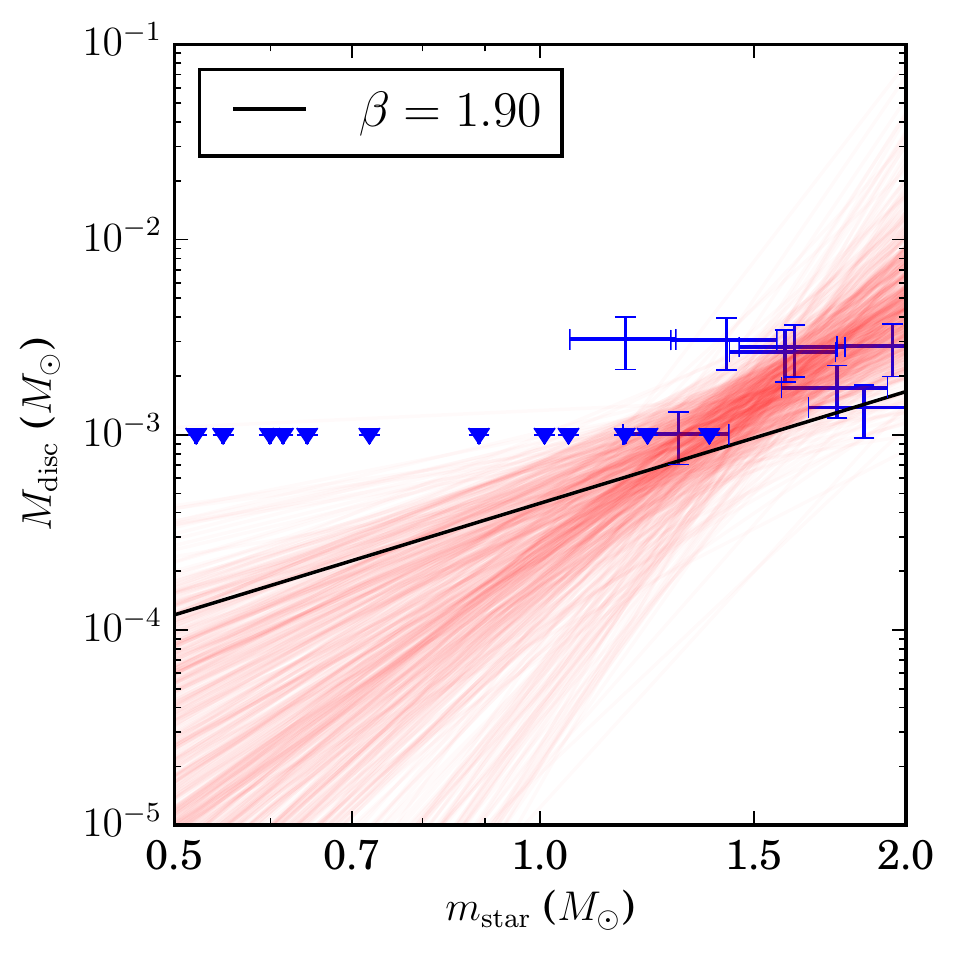}
     \caption{{A subset of 20 discs of those shown in figure} \ref{subfig:mstvmdisc} {(after $3$~Myr of evolution), chosen to be approximately uniformly distributed over a range of stellar masses $0.5-2\, M_\odot$. All discs are also selected such that they are in an environment where the projected flux has an instaneous value $3000 \, G_0 < F_\mathrm{FUV}< 8000 \, G_0$. Discs with $M_\mathrm{disc}<10^{-3}\, M_\odot$ are considered upper limits (non-detections). The red lines are a subset of samples from the posterior distribution obtained from Markov chain Monte Carlo modelling} \citep[using the \textsc{Linmix} package -][]{Kel07}. {The black line is  is a model with $\beta =1.9$ selected from the posterior distribution such that the true $\beta$ is greater than this value with $2\sigma$ confidence.}}
     \label{fig:mstvdisc_obs}
   \end{figure}  
   
  {The relationship between final disc mass and stellar mass is considered in figure} \ref{subfig:mstvmdisc} (figure \ref{subfig:mstvmdisc_dist}) {for disc masses initially correlated (uncorrelated) with} the host star mass. {We obtain power law indices ($M_\mathrm{disc} \propto m_\mathrm{star}^\beta$) $\beta = 3.93\pm 0.11$ and $2.75 \pm 0.12$ respectively. In both cases this is a substantially more superlinear relationship than in local PPD populations that have not been significantly photoevaporated}  \citep[$1 < \beta <1.9$][]{Andr13, Pas16}. Physically this is because $\dot{M}_\mathrm{wind}$ is strongly dependent on stellar mass such that discs around low mass stars are depleted much faster (due to a shallower potential) than those around high mass stars for a fixed flux and disc radius \citep{Haw18b}. In regions where a disc population has undergone significant photoevaporation we then expect lower mass stars to host lower mass PPDs. {Even if it proves impossible to obtain sufficiently large samples of mm-based mass determinations in Cyg OB2 in order to conduct the comparison made in figure} \ref{fig:cumfracs}, an alternative way to test the role of external photoevaporation would be to examine any evidence for steep disc mass--stellar mass relationship. Indeed, \citet{Ans17} find a steepening of the relationship with age across different regions that could be due to external photoevaporation. To the contrary, \citet{Eis18} find a shallow disc mass--host mass relationship in the strongly irradiated discs of the Orion Nebula Cluster; this could be the result of the youth or complex formation history of the stellar population \citep{Bec17, Kro18}. Disentangling these effects requires a detailed modelling of those regions, such as that presented in this paper.
  
 {To estimate whether detecting large $\beta$ values is possible in Cyg OB2, we extract a subset of $20$ discs from our model, choosing them such that they lie in a region of projected FUV flux $3000 \, G_0 < F_\mathrm{FUV} < 8000\, G_0$. A lower limit of $3000 \, G_0$ is appropriate since below this figures} \ref{subfig:g0bar} and \ref{fig:g0bart_dist} suggest that alternative processes to external photoevaporation have a comparable influence on disc evolution. We also select stellar masses to be approximately uniformly distributed between $0.5$ -- $2 \, M_\odot$. We assume error of $30\%$ in $M_\mathrm{disc}$, with $M_\mathrm{disc}< 10^{-3}\, M_\odot$ constituting a non-detection, and a $10\%$ error in $M_\mathrm{star}$. We then attempt to fit a power law using the \textsc{Linmix} package \citep{Kel07} {for a number of different PPD subsets. An illustrative example of such an exploration is shown in figure} \ref{fig:mstvdisc_obs}. {In all cases we find that the fit effectively discounts power law relations with $\beta \leq 1.9$. {Values of $\beta \leq 1.9$ are similarly discounted in the model with stellar host mass independent disc initial conditions. Our findings suggest that, if a sample is carefully selected, evidence of external photoevaporation should be detectable in future observations of Cyg OB2.}
  
 One might ask whether it is possible to find signatures of underlying substructure within a population of PPDs. Given that the dynamical history is linked to the irradiation by FUV photons, it may be possible to detect distinct groups in the mass distribution of a disc sample. As discussed previously, in an externally photoevaporated population, disc properties are more strongly correlated with host mass than FUV flux. Therefore considering disc properties as a function of host mass is the best chance of finding distinct groups within a sample. This possibility is explored in figure \ref{fig:mstvdisc}, where the largest scale fractal membership is indicated for each PPD. We find  that the mass distribution does not demonstrate a clear segregation between groups. {The subgroups are in themselves massive ($\sim 10^4 \, M_\odot$) and the high mass end of the IMF is therefore well sampled in each filament. It is possible that for smaller scale substructure (with filaments of mass $<10^3 \, M_\odot$, see} \citealt{Win18b}) {a less well sampled IMF might mean that different filaments have quite different local $F_\mathrm{FUV}$. In this case we should expect to see a distinct PPD mass and outer radius distributions between filaments. }

\subsubsection{Disc radii}

  \begin{figure*}
     \subfloat[\label{subfig:mstvrdisc} ] { \includegraphics[width=0.5\textwidth]{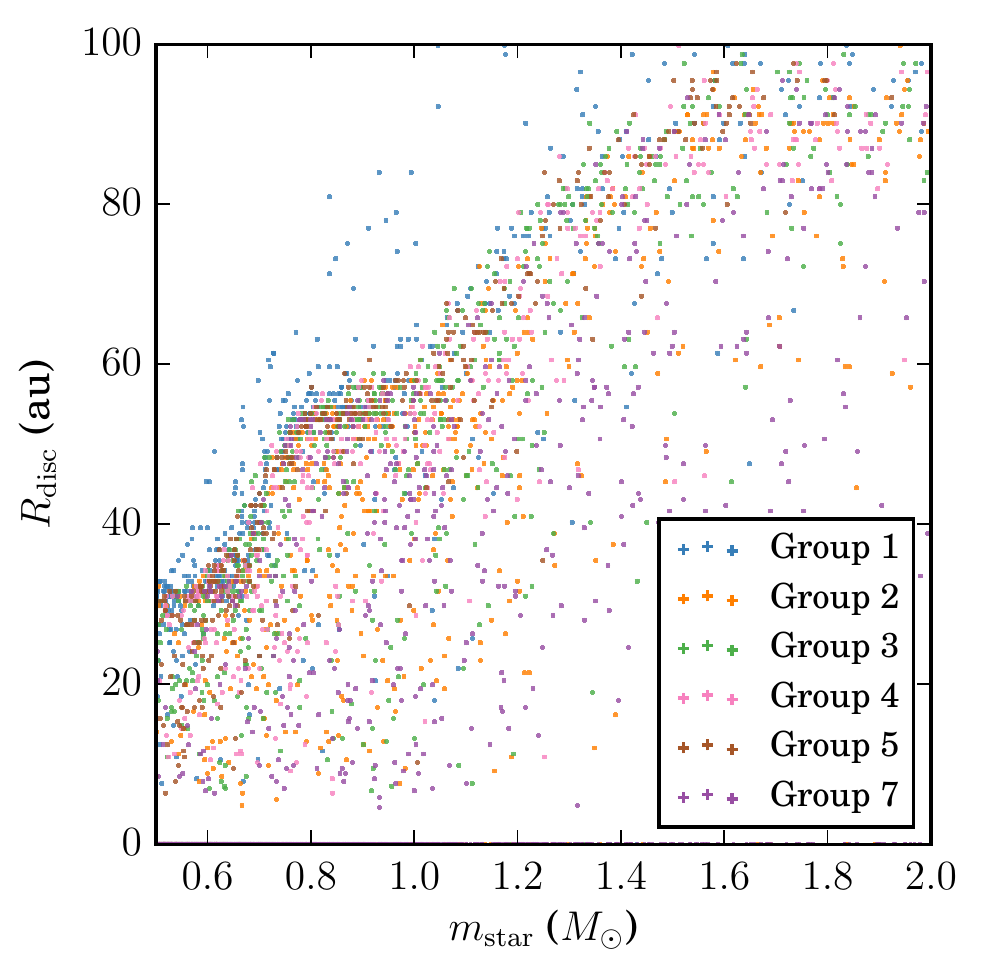}
     }
     \subfloat[\label{subfig:mstvrdisc_dist} ] { \includegraphics[width=0.5\textwidth]{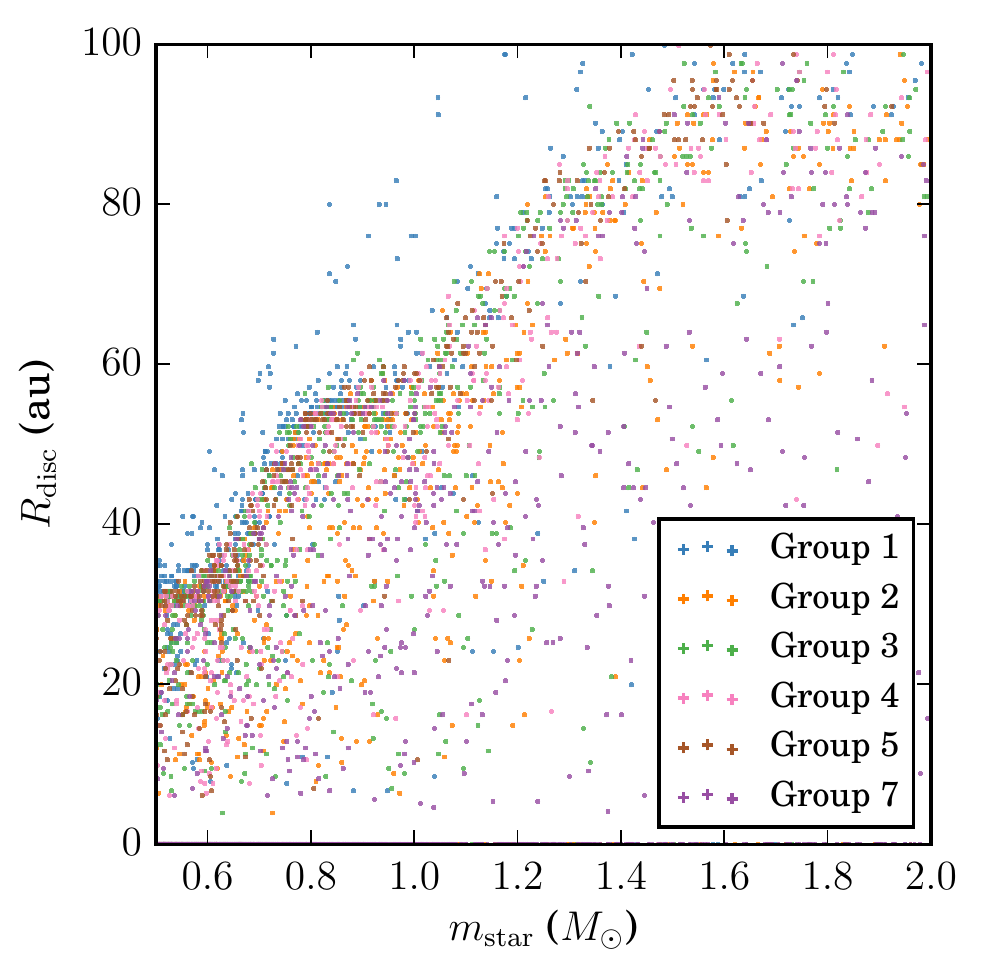}
     }
     \caption{As in figure \ref{fig:mstvdisc} but for disc outer radius distributions. In figure \ref{subfig:mstvrdisc} initial disc masses are drawn from a distribution which scales linearly with stellar host mass, while in figure \ref{subfig:mstvrdisc_dist} the initial PPD mass is not correlated to host mass. Points are coloured by the largest scale fractal membership. We find that the outer radii are largely independent of disc initial conditions and are correlated with stellar host mass.}
     \label{fig:mstvrdisc}
   \end{figure*}

An alternative observable property, the outer disc radius $R_\mathrm{out}$, may prove to be a better probe of substructure in future observations. This is because, while $M_\mathrm{disc}$ is depleted over the entire lifetime of the disc, the outer radius in regions of strong FUV flux is set by a balance between photoevaporative mass loss and viscous expansion on short timescales. We have so far neglected discussion of disc radius simply because resolving disc radii down to $\sim 10$s~au is challenging in a region such as Cyg OB2; the highest resolution of \textit{ALMA} is $\sim 0.02''$, corresponding to $\sim 30$~au at $1.33$~kpc distance. However, because it may be possible to resolve disc radii to sufficient accuracies in closer young stellar environments, we illustrate the distribution of disc radii in figure \ref{fig:mstvrdisc}. There are some indications of different radius distributions between different fractal associations, but no clear segregation between them. The majority of discs have $R_\mathrm{disc} < 60$~au  which would require $< 0.05 ''$ resolution at the distance of Cyg OB2. {We note that outer disc radius measurements can be subject to large uncertainties, and that the dust emission is frequently less extended than the gas \citep[e.g.][]{Bir10, Gui11, deG13}. This discrepancy may be less problematic in regions where disc radii are externally suppressed \citep[for example close stellar encounters may result in wavelength independent PPD outer radius measurements, as in the case of the disc around HV Tau C - ][]{Mon00,Win18c}. This assumption requires further study into gas-dust physics in externally photoevaporating discs (see \citealt{Cle16} for discussion of gas and dust evolution in an FUV irradiated PPD). {Overall we see the same trend as in disc masses that the outer radius is correlated with host mass. This is again because $\dot{M}_\mathrm{wind}$ decreases with $m_\mathrm{star}$, such that the radius at which viscous expansion is in equilibrium with FUV induced mass loss is more extended for more massive host stars. }

\section{Conclusions}
\label{sec:concs}

Using $N$-body simulations and viscous disc evolution models we have successfully reproduced the properties of Cygnus OB2, including the stellar kinematics and the surviving PPD fractions as a function of projected FUV flux. Our modelling supports the following scenario:
\begin{itemize}
	\item {If the viscous evolution of PPDs is well described by a} \mbox{\citeauthor{Sha73}} $\alpha$-parameter $\alpha = 10^{-2}$ ($\tau_\mathrm{visc} \approx 0.5$~Myr for a solar mass star), then expulsion of the primordial gas content in the region must have concluded $\sim 0.5$~Myr ago. Approximately $0.5$~Myr of exposure to strong FUV fields is required to reproduce the current disc survival rates as a function of (projected) flux. This value of $\alpha$ will be an upper limit if FUV extinction due to the primordial gas was not efficient (because of a clumpy spatial distribution, for example). 
	\item The initial three dimensional velocity dispersion must have been $\sim 50$~km/s in order to be consistent with the present day central velocity dispersion. This is large with respect to observed stellar populations, even for OB associations (e.g. \mbox{\citealt{War18}} and references therein). However, velocity gradients $\sim 10$~kms$^{-1}$pc$^{-1}$ are found across giant molecular clouds, particularly in the case of cloud-cloud collision \citep{Wu15, Bis18, Pol18}. 
	\item Anisotropy in the present day velocity dispersion requires the presence of primordial subtructure. We find that Cyg OB2 must have been comprised of large scale filaments (or fractal clumps), with a mass $\sim 10^4$~$M_\odot$.
	\item In such a model, a gas content of $\sim 8 \times 10^6\, M_\odot$ is required to maintain initial virial equilibrium. This is massive compared to the known distribution of molecular cloud masses in the Milky Way \mbox{\citep{Lon14}}. 
	\item The total stellar mass required to sustain a sufficient central mass after $3$~Myr is $\sim 8 \times 10^4$~$M_\odot$. This suggests a star formation efficiency of $\sim 1 \%$. 
	\item The apparent lack of expansion measured by \mbox{\citet{Wri16}} {need not be interpreted as evidence of no physical expansion of the stellar population. We find that our (expanding) model is equally consistent with an expansion parameter $\mathcal{E} \approx 0.5$. Such a measurement can only be interpreted probabilistically for an association with large scale substructure.}
\end{itemize} 

{The primary caveat of these conclusions is that the gas expulsion timescale derived by disc fractions is degenerate with $\alpha$. Our choice ($\alpha = 10^{-2}$, although this is in turn dependent on choice of scale radius $R_1$ - see equation} \ref{eq:tvisc}) {yields a good fit to the disc fractions and also allows us to reproduce kinematic observations with gas expulsion timescale $\tau_\mathrm{gas} = 2.5$~Myr. In future, observations of PPD populations in regions with a strong FUV flux but with more modest velocity dispersions (and primordial gas density) might offer further constraints for $\alpha$.}

Finally we make predictions for future observations of the disc mass and radius distribution in Cyg OB2. We find that samples of PPDs in highly FUV irradiated environment have significantly reduced masses and outer radii than in regions of more modest flux. {However, taking into account sensitivity limits, statistically observing these differences requires sample sizes of $\gtrsim 100$s of discs in both high and low FUV flux bins. This is similarly true for disc outer radii, and makes probing the difference between disc property distributions as a function of FUV environment directly impractical at present.}

{In Cygnus OB2 we expect a strong correlation between stellar mass and disc mass ($M_\mathrm{disc} \propto m_\mathrm{star}^{\beta}$, with $\beta >2$) which is a consequence of the fact that, for a disc with given radius and in a given FUV environment, the mass loss rate is higher in the shallower potential of low mass stars. This effect should be clear in a sufficiently large sample of irradiated discs in any environment. In Cygnus OB2, we demonstrate that - taking into account the finite sensitivity of \textit{ALMA} - it would be sufficient to target around $20$ stars of known mass in order to demonstrate a value of $\beta$ that is steeper than the canonical value $\sim 1$ -- $1.9$ which is seen in non-irradiated disc samples.} We conclude that, pending empirical confirmation, discs around low mass stars born in environments of strong FUV flux are likely to have a significantly depleted mass budget for planet formation. {Indeed, unless dust within PPDs can rapidly grow to size scales where it is immune to photoevaporative stripping} \mbox{\citep[e.g.][]{You05}}, planet formation may be completely supressed for such host stars \mbox{\citep{Haw18}}.  

\section*{Acknowledgements}

We would like to thank Tom Haworth for providing access to the FRIED grid of mass loss rates ahead of publication. AJW thanks Sverre Aarseth and Long Wang for helpful discussion regarding $N$-body simulations. We thank the anonymous referee for a considered report which improved the clarity of the manuscript. This work has been supported by the DISCSIM project, grant agreement 341137 funded by the European Research Council under ERC-2013-ADG. AJW thanks  the  Science  and  Technology  Facilities  Council  (STFC)  for  their  studentship.




\bibliographystyle{mnras}
\bibliography{truncation} 



\bsp	
\label{lastpage}
\end{document}